\documentclass[prl,twocolumn,preprintnumbers,amsmath,amssymb,superscriptaddress,linenumbers,longbibliography]{revtex4-1}

\usepackage{graphicx}
\usepackage{dcolumn}
\usepackage{bm}
\usepackage{float}
\usepackage{hyphenat}
\usepackage[dvipsnames]{xcolor}
\usepackage{physics}
\usepackage[T1]{fontenc}

\newcommand{\heriotwatt}{Institute of Photonics and Quantum Sciences, SUPA, Heriot-Watt University, Edinburgh EH14 4AS, UK}
\newcommand{\TsukubaKenji}{Research Center for Functional Materials, National Institute for Materials Science, 1-1 Namiki, Tsukuba 305-0044, Japan}
\newcommand{\TsukubaTakashi}{International Center for Materials Nanoarchitectonics, National Institute for Materials Science,  1-1 Namiki, Tsukuba 305-0044, Japan}
\newcommand{\Imperial}{Departments of Materials and Physics and the Thomas Young Centre for Theory and Simulation of Materials, Imperial College London, South Kensington Campus, London SW7 2AZ, United Kingdom}

\newcommand{\Trieste}{Dipartimento di Fisica, Universit\'a di Trieste, strada Costiera 11, 34151, Trieste, Italy}

\begin{document}
\nolinenumbers
\title{The interplay of field-tunable strongly correlated states in a multi-orbital moir{\'e} system}

\author{Aidan J. Campbell}
\affiliation{\heriotwatt}
\author{Valerio Vitale}
\affiliation{\Trieste}
\affiliation{\Imperial}
\author{Mauro Brotons-Gisbert}
\affiliation{\heriotwatt}
\author{Hyeonjun Baek}
\affiliation{\heriotwatt}
\author{Takashi Taniguchi}
\affiliation{\TsukubaTakashi}
\author{Kenji Watanabe}
\affiliation{\TsukubaKenji}
\author{Johannes Lischner}
\affiliation{\Imperial}
\author{Brian D. Gerardot}
\email{B.D.Gerardot@hw.ac.uk}
\affiliation{\heriotwatt}
    
\date{\today}
\begin{abstract}

The interplay of charge, spin, lattice, and orbital degrees of freedom leads to a wide range of emergent phenomena in strongly correlated systems. In heterobilayer transition metal dichalcogenide moir{\'e} systems, recent observations of Mott insulators and generalized Wigner crystals are well described by triangular lattice single-orbital Hubbard models based on K-valley derived moir{\'e} bands. Richer phase diagrams, mapped onto multi-orbital Hubbard models, are possible with hexagonal lattices in $\Gamma$-valley derived moir{\'e} bands and additional layer degrees of freedom. Here we report the tunable interaction between strongly correlated hole states hosted by $\Gamma$- and K-derived moir{\'e} bands in a monolayer MoSe$_2$ / natural WSe$_2$ bilayer device. To precisely probe the nature of the correlated states, we optically characterise the behaviour of exciton-polarons and distinguish the layer and valley degrees of freedom. We find that the honeycomb $\Gamma$-band gives rise to a charge-transfer insulator described by a two-orbital Hubbard model with inequivalent $\Gamma_\mathrm{A}$ and $\Gamma_\mathrm{B}$ orbitals. With an out-of-plane electric field, we re-order the $\Gamma_\mathrm{B}$- and K-derived bands energetically, driving an abrupt redistribution of carriers to the layer-polarized K orbital where new correlated states are observed. Finally, by fine-tuning the band-alignment, we obtain degeneracy of the $\Gamma_\mathrm{B}$ and K orbitals at the Fermi level. In this critical condition, stable Wigner crystals with carriers distributed across the two orbitals are observed until the Fermi-level reaches one hole per lattice site, whereupon the system collapses into a filled $\Gamma_\mathrm{B}$ orbital. Our results establish a platform to investigate the interplay of charge, spin, lattice, and layer geometry in multi-orbital Hubbard model Hamiltonians. 

\end{abstract}

\maketitle

\section{Introduction}

Although conceptually simple, the single-orbital two-dimensional Hubbard model can describe many ingredients of strongly correlated systems. However, real quantum materials are typically described using multi-orbital Hubbard models. For instance, doping in the two-orbital charge-transfer insulator has been connected to high-temperature superconductivity in the cuprates \cite{emery1987theory,zhang1988effective,dagotto1994correlated} and iron-based superconductors demand Hubbard models that account for multiple orbitals degenerate at the Fermi level \cite{dai2012magnetism}. While significant challenges exist to accurately solve the single-orbital Hubbard model and its derivatives, the phase diagrams of more complex multi-orbital models become even more intractable to elucidate; they sensitively depend on the interplay of the different degrees of freedom \cite{rohringer2018diagrammatic}. Hence, there is a strong motivation to develop platforms for strongly correlated physics which are relatively simple yet possess a high degree of tunability such that the emergent physics can be probed and accurately mapped onto paradigmatic quantum Hamiltonians \cite{gross2017quantum,kennes2021moire}.    

Synthetic moir{\'e} superlattices, formed by stacking two atomic layers with a lattice mismatch and / or relative twist angle, have emerged as a versatile solid-state platform that potentially fulfils these criteria  \cite{kennes2021moire,balents2020superconductivity,mak2022semiconductor}. The versatility of moir{\'e} materials arises from the impressive array of tuning knobs that can effectively control the moir{\'e} lattice and adjust the strength and nature of the particle interactions. For instance, electrostatic gating can fine-tune the fractional filling ($\nu$) of the lattice with carriers (electrons or holes, $\nu_e$ or $\nu_h$, respectively), displacement fields can adjust the moir{\'e} potential depth \cite{ghiotto2021quantum,li2021continuous}, and the long-range Coulomb interactions can be manipulated with screening gates \cite{stepanov2020untying, li2021charge}. Further, in semiconducting transition metal dichalcogenide (TMD) moir{\'e} systems, the Coulomb interaction strengths between particles in the moir{\'e} lattice can be continuously tuned via the moir{\'e} period \cite{wu2018hubbard,zhang2020moire}. Additionally, different real-space lattice geometries can be found in the low energy valence moir{\'e} bands, dependent on the valley degree of freedom: moir{\'e} orbitals on a triangular lattice are generally found for K-valley moir{\'e} bands localized on a single atomic layer \cite{wu2018hubbard,zhang2020moire,tang2020simulation,regan2020mott,shimazaki2020strongly,wang2020correlated,campbell2022strongly,xu2022tunable} while honeycomb lattices are found for $\Gamma$-valley moir{\'e} bands spread across each layer \cite{angeli2021gamma,xian2021realization,kaushal2022magnetic,an2020interaction,pei2022observation,gatti2022observation,foutty2022tunable}. Notably, the energetic ordering of $\Gamma$- versus K-derived valence moir{\'e} bands is sensitively dependent on material combination, twist angle, and atomic relaxation effects \cite{vitale2021flat}. To date, most experimental observations of correlated states in both TMD moir{\'e} homo- and hetero-bilayers have been sufficiently described by single-band Hubbard models. Thus, experimentally realizing tunable moir{\'e} systems that portray multi-orbital Hubbard physics remains an exciting opportunity in this emerging platform. 

Here, by adding a non-twisted monolayer to a moir{\'e} hetero-bilayer (HBL) device to create a hetero-trilayer (HTL) device with only one moir{\'e} interface, we demonstrate a new degree of freedom to realize an extended $t$-$U$-$\mathcal{V}$ three-orbital Hubbard system based on low energy moir{\'e} bands at $\Gamma$ and K which host distinct real-space lattice geometries. Crucially, the relative ordering of these bands can be fine-tuned with an external electric (displacement) field $\vec{E}$ and we are able to investigate the interplay of correlated states between different orbitals. To precisely probe the nature of the correlated states, we develop an optical spectroscopy technique based on the distinctive behaviour of exciton-polarons \cite{back2017giant,roch2019spin,liu2021exciton} that unambiguously distinguishes the layer and valley degrees of freedom (see Methods).

\section{Layer and valley degrees of freedom for holes in moir{\'e} bands}

\begin{figure*}
    	\begin{center}
    		\includegraphics[scale= 0.4]{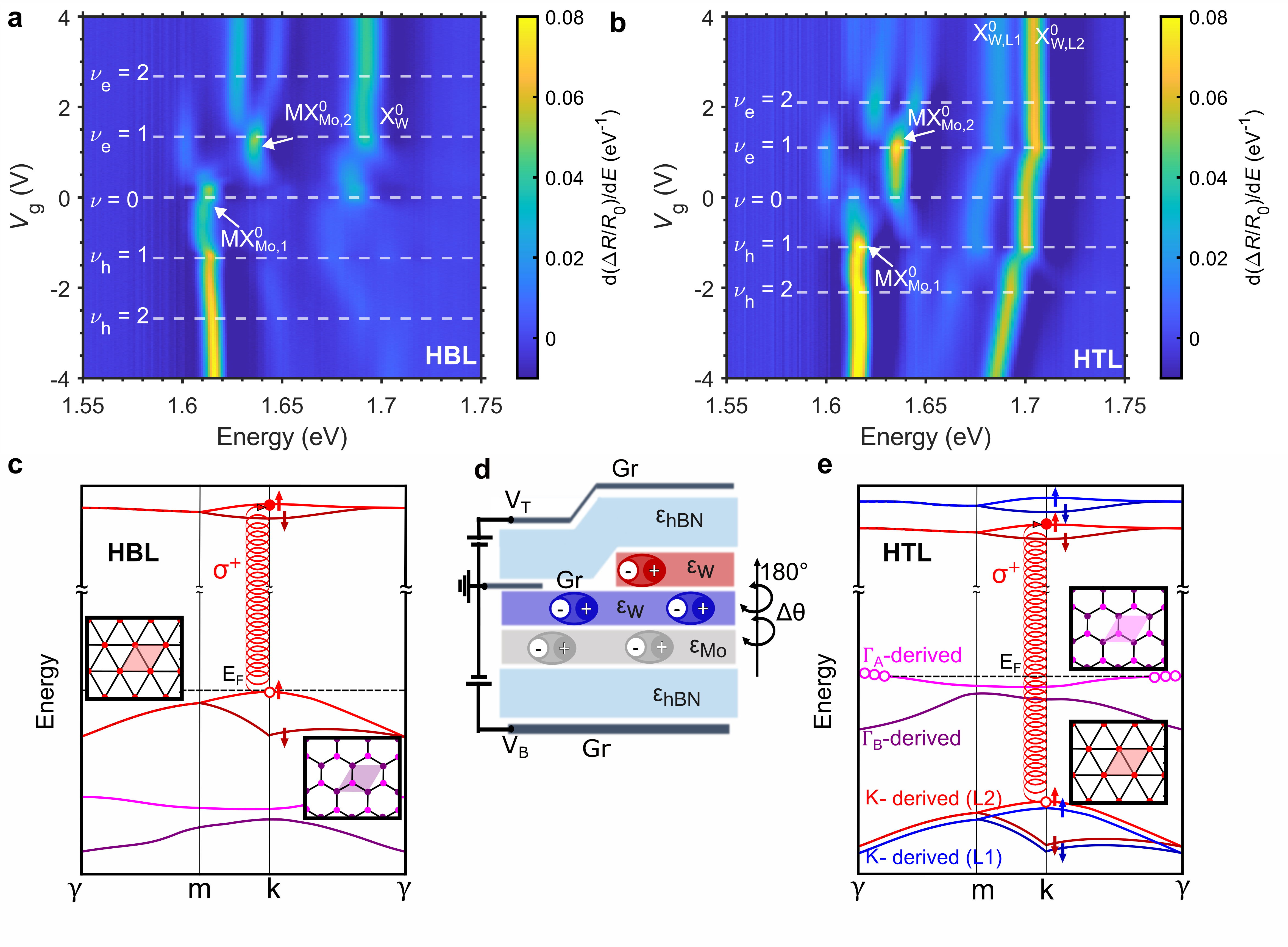}
    	\end{center}
        \caption{\textbf{Layer and valley properties of the excitons and holes in a moir{\'e} heterotrilayer device.}  \textbf{a,b,} Density plots of the $V_\mathrm{g}$ dependence of d($\Delta R/R_0$)/d$E$) in the (a) 1L WSe$_2$/1L MoSe$_2$ (HBL) and (b) 2L WSe$_2$/1L MoSe$_2$ (HTL)  heterostructure regions. The MoSe$_2$ moir{\'e} excitons, MX$_\mathrm{Mo,1}$ and MX$_\mathrm{Mo,2}$, and WSe$_2$ neutral excitons, X$^0_\mathrm{W}$, are indicated. In the HTL, the intralayer excitons localised in the interface (L1) and upper (L2) layers of the bilayer WSe$_2$ are non-degenerate due to different dielectric environments ($\epsilon$), as indicated in panel (d). The white dashed lines indicate integer values of electrons and holes per moir{\'e} unit cell, $\nu_e$ and $\nu_h$, respectively. \textbf{d} Schematic of the full device with HBL and HTL regions. The TMD layers are fully encapsulated in hexagonal boron nitride (hBN). Graphene (Gr) layers form contacts to the top and bottom hBN and the heterostructure. \textbf{c,e} Band structure schematics for the moir{\'e} Brillouin zone based on DFT calculations in the (c) HBL and (e) HTL regions. In the HBL, optically injected excitons reside in $\pm$K-derived states and are dressed solely by holes in $\pm$K moir{\'e} orbitals. In the HTL, the excitons are dressed by holes at both $\Gamma$ and $\pm$K moir{\'e} bands. 
        As indicated in the insets, the $\Gamma$ moir{\'e} orbitals form a honeycomb lattice with inequivalent A and B sites. The moir{\'e} orbitals from the $\pm$K valleys form a triangular lattice.}
        \label{fig:1}
\end{figure*}
Our HTL sample is a natural bilayer (2L) 2$H$-WSe$_2$ / monolayer (1L) MoSe$_2$ device with a moir{\'e} interface period of $\approx$ 7 nm due to a relative twist of $\theta \approx 2.7^\circ$. Compared to a 1L WSe$_2$ / 1L MoSe$_2$ HBL device, the HTL provides an additional layer degree of freedom \cite{shimazaki2020strongly,xu2022tunable,zhou2021bilayer} while also raising the prospect for exploitation of the valley degree of freedom in the valence band \cite{movva2018tunable}. To aid unambiguous identification of the spectral features in the HTL device, we investigate a sample (see Fig.~\ref{fig:1}d for a cross-section schematic of the device) with adjacent HBL and HTL regions with nominally identical moir{\'e} period at each  WSe$_2$ / MoSe$_2$ interface (see Methods).  Hence, the HBL serves as a reference with well understood exciton-polaron behavior in the presence of strongly correlated states \cite{campbell2022strongly}. 

Figure~\ref{fig:1}a,b shows doping ($V_\mathrm{g}$)-dependent sweeps (see Methods) while monitoring the derivative of the reflection contrast with respect to energy (d($\Delta R/R_0$)/d$E$) in representative HBL and HTL regions of the device, respectively, at a temperature of 4K. We observe two unequivocal signatures of the effect of the moir{\'e} potential in both regions. First, at charge neutrality, we observe two MoSe$_2$ excitonic features, labelled MX$_\mathrm{Mo,1}$ and MX$_\mathrm{Mo,2}$, which we assign to be MoSe$_2$ moir{\'e} intralayer excitons, in agreement with recent reports \cite{jin2019observation,ruiz2019interlayer,campbell2022strongly}. Second, we observe doping-dependent periodic modulations in the peak position, intensity, and linewidth of all intralayer excitonic features at $V_\mathrm{g}$ values corresponding to integer fractional fillings of the moir{\'e} superlattice (white dashed lines in Fig.~\ref{fig:1}a,b), indicating the formation of correlated electron and hole states. In the HTL, the $V_\mathrm{g}$ required to fill the moir{\'e} lattice with either one electron or one hole per site is similar to the HBL and corresponds to a twist angle of 2.7$^\circ$ (see Methods). 

In contrast to the HBL region, in the HTL we observe two WSe$_2$ intralayer exciton peaks with different transition energies due to their different dielectric environments (see sketch in Fig.~\ref{fig:1}d). The lower energy feature, labelled X$^0_\mathrm{W,L1}$, is the WSe$_2$ neutral exciton localised in the heterostructure interface layer (L1), while the higher energy exciton, X$^0_\mathrm{W,L2}$, is localised in the other layer (L2) (see Methods). This spectroscopic signature provides a clear probe to determine which WSe$_2$ layer is occupied by the Fermi sea under doping (via the formation of attractive exciton-polarons for each intralayer exciton). We measure multiple spatial locations in the HTL region and find the behaviour shown in the main text is general (see Suppl. Fig. S1 and Suppl. Tab. S1). 

Figure~\ref{fig:1}c,e shows the valence band schematics in the moir\'e Brillouin zone for the HBL and HTL, respectively, based on calculations using ab initio density functional theory (DFT) (see Suppl. Sec. S2). Moir{\'e} bands derive from either states near the $\Gamma$ point or the $\pm$K points. Doped carriers that occupy the moir{\'e} valence band arising from $\pm$K form a two-dimensional triangular lattice. In contrast, the holes in the moir{\'e} band derived from $\Gamma$ form a honeycomb lattice with inequivalent A and B sites (see Fig.~\ref{fig:1}e). In addition, the moir{\'e} bands inherit the contrasting properties of the different Brillouin zone locations. Holes in the K-derived bands are highly localised in one of the WSe$_2$ layers as shown by the layer distribution of the hole wavefunction (see Suppl. Fig. S9) while $\Gamma$ holes are delocalised across both WSe$_2$ layers. The $\Gamma_\mathrm{A}$ band is the lowest energy available valence moir{\'e} band. 
In each sample, optically excited electron-hole pairs reside in $\pm$K-derived bands, while carriers that dress the excitons are doped into the lowest energy available valence band, which can be either $\pm$K-derived or $\Gamma$-derived. According to the selection rules, for the HTL, $\sigma^+$ ($\sigma^-$) light probes the $+$K ($-$K) derived states in L1 and the $-$K ($+$K) derived states in L2 of the natural bilayer WSe$_2$, due to the 2$H$ stacking and spin-layer locking \cite{jones2014spin,brotons2020spin}.



\begin{figure*}
    	\begin{center}
    		\includegraphics[scale= 0.45]{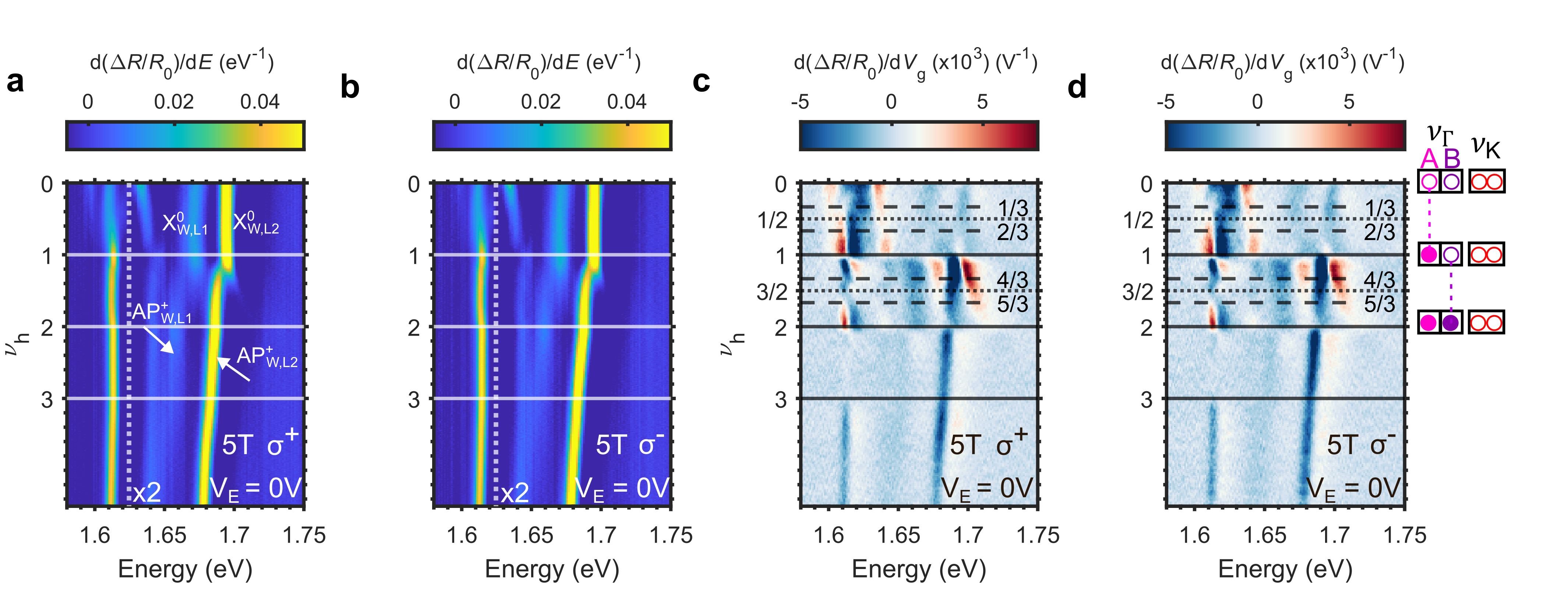}
    	\end{center}
        \caption{\textbf{$\Gamma$-valley correlated hole states in the moir{\'e} heterotrilayer.} \textbf{a} $\sigma^+$-  and \textbf{b} $\sigma^-$-helicity-resolved  evolution of d$(\Delta R/R_0)$/d$E$, as a function of $\nu_h$ under an applied magnetic field of 5~T. \textbf{c,d} Density plot of the first derivative with respect to gate voltage of $\Delta R/R_0$ (d($\Delta R/R_0$)/d$V_\mathrm{g}$) in panel (a) and (b) respectively. The orbital at the Fermi level is depicted in the schematic to the right. On all plots, the solid lines indicate integer hole fractional fillings of the moir{\'e} superlattice, while the dashed and dotted lines in (c,d) show intermediate fractional fillings.}
        \label{fig:2}
\end{figure*}

\section{Probing $\Gamma$ valley correlated electronic states via exciton-polaron spectroscopy}

Figure~\ref{fig:2}a,b shows the $\sigma^+$- and $\sigma^-$-resolved $\nu_h$  dependence of the d($\Delta R/R_0)$/d$E$ signal, respectively, in the HTL under application of 5~T magnetic ($B$) field. The solid lines indicate integer values of $\nu_h$ (i.e. the total number of holes per moir{\'e} unit cell). To further highlight the formation of correlated electronic states, we plot the derivative of the reflection contrast with respect to $V_\mathrm{g}$, d($\Delta R/R_0$)/d$V_\mathrm{g}$, as shown in Fig.~\ref{fig:2}c,d. 
We note the largest changes in d($\Delta R/R_0$)/d$V_\mathrm{g}$ (at $V_\mathrm{g} = -1.1$~V, $-2.1$~V, and $-3.1$~V) identify the formation of different insulating correlated hole states.  Consistent with previous works \cite{tang2020simulation,campbell2022strongly,li2021continuous}, we use these $V_\mathrm{g}$ values to define the number of holes per moir{\'e} unit cell, $\nu_h = 1, 2, 3$. We also observe changes in d($\Delta R/R_0$)/d$V_\mathrm{g}$ at further intermediate hole fillings (black dashed lines in Fig.~\ref{fig:2}c,d), which we ascribe to the formation of generalised Wigner crystals. We perform temperature-dependent sweeps of $V_\mathrm{g}$ while monitoring $\Delta R/R_0$ and find a critical temperature $T_c \sim 98$~K for the correlated state at $\nu_h = 1$ and $T_c \sim$ 59~K for the correlated state at $\nu_h = 2$ (see Suppl. Figs. S4 and S5). The $V_\mathrm{g}$ required to fill the moir{\'e} lattice with either one electron or one hole per site is identical and corresponds to a twist angle of 2.7$^\circ$ (see Methods).

Next, we observe that there is little change in the two WSe$_2$ excitonic resonance energies as the doping is increased from zero to $\nu_h = 1$. After $\nu_h = 1$, X$^0_\mathrm{W,L1}$ and X$^0_\mathrm{W,L2}$ abruptly red-shift, indicating the formation of the hole-dressed attractive polarons AP$^+_\mathrm{W,L1}$ and AP$^+_\mathrm{W,L2}$ \cite{courtade2017charged,liu2021exciton,wang2018electrical}. Two key pieces of experimental evidence show the valence band maximum in the system derives from the $\Gamma$ valley, as predicted by the DFT calculations (see Fig.~\ref{fig:1}e). First, we observe that AP$^+_\mathrm{W,L1}$ and AP$^+_\mathrm{W,L2}$ \emph{red-shift} with increasing $\nu_h$. This is in contrast to natural $2H$~bilayer WSe$_2$, where an overall blue-shift of the AP$^+_W$ occurs due to Pauli blocking effects: resident carriers fill up states at the valence band at the $\pm$K point where the photo-excited hole also resides \cite{wang2018electrical} (see Suppl. Sec. S1). The contrasting red-shift in the moir{\'e} HTL implies that the valence band maximum is \emph{not} derived from $\pm$K states. Second, unlike the examples of strong spin polarization of holes residing in $\pm$K valleys \cite{back2017giant,roch2019spin}, in Fig.~\ref{fig:2}a,b we observe negligible differences in the intensity and energy of AP$^+_\mathrm{W,L1}$ and AP$^+_\mathrm{W,L2}$ between the $\sigma^+$- and $\sigma^-$-resolved sweeps at an applied magnetic field of 5~T, indicating an absence of a spin-polarisation for the doped holes. This finding is consistent with a photo-excited exciton which resides at $\pm$K dressed by doped holes in moir{\'e} bands derived from the $\Gamma$ point, in agreement with DFT results. We remark that the formation of AP$^+_\mathrm{W,L1}$ and AP$^+_\mathrm{W,L2}$ at the identical $V_\mathrm{g}$ value is further evidence of a $\Gamma$-derived lowest energy valence band: holes in the $\Gamma$ band are delocalised across the two WSe$_2$ layers and dress both X$^0_\mathrm{W,L1}$ and X$^0_\mathrm{W,L2}$ equally. In the absence of delocalisation of holes, we would expect carriers to sequentially fill the L1 and L2 derived moir{\'e} bands, causing a $V_\mathrm{g}$ offset in the formation of AP$^+_\mathrm{W,L1}$ and AP$^+_\mathrm{W,L2}$. 

We now discuss the nature of the correlated state that arises at $\nu_h = 1$  in the $\Gamma$ orbital, which is determined to be a charge transfer insulator. The energy gap between the two $\Gamma$ moir{\'e} bands that map onto inequivalent A and B sites of a honeycomb lattice ($\Gamma_A$ and $\Gamma_B$ orbitals, respectively) is predicted by DFT to be $\sim 18$ meV (see Supp. Sec. S2), comparable to the value of the charge gap we can tentatively estimate from the temperature at which the correlated state at $\nu_h= 1$ disappears, which is $\sim 10$ meV (see Suppl. Fig. S5). Therefore, the $\Gamma_\mathrm{A}$ lattice is first completely filled, followed by sequential filling of the $\Gamma_\mathrm{B}$ sites. In contrast, for a Mott state, for $1<\nu_h<2$ carriers would continue filling the A orbital, leading to doubly occupied sites. The A-B sublattice inequivalency also presents a likely explanation for the lack of formation of AP$^+_\mathrm{W,L1}$ and AP$^+_\mathrm{W,L2}$ up to $\nu_h=1$. The DFT calculations reveal that the photo-excited hole at K resides on the B sublattice site, while for $0<\nu_h<1$ the doped hole in the $\Gamma$ moir{\'e} band lies on the A site (see Suppl. Fig. S9). Hence, there is no spatial or momentum overlap between the two states, minimizing the formation of an attractive exciton-polaron \cite{naik2022nature}. For $1<\nu_h<2$, holes are doped into the $\Gamma_\mathrm{B}$ moir{\'e} orbital, which spatially overlaps with the photo-excited hole. Hence, the attractive polarons form in each layer for this range of fractional filling.

\section{Electric field tuning of moir{\'e} orbitals}

\begin{figure*}
    	\begin{center}
    		\includegraphics[scale= 0.45]{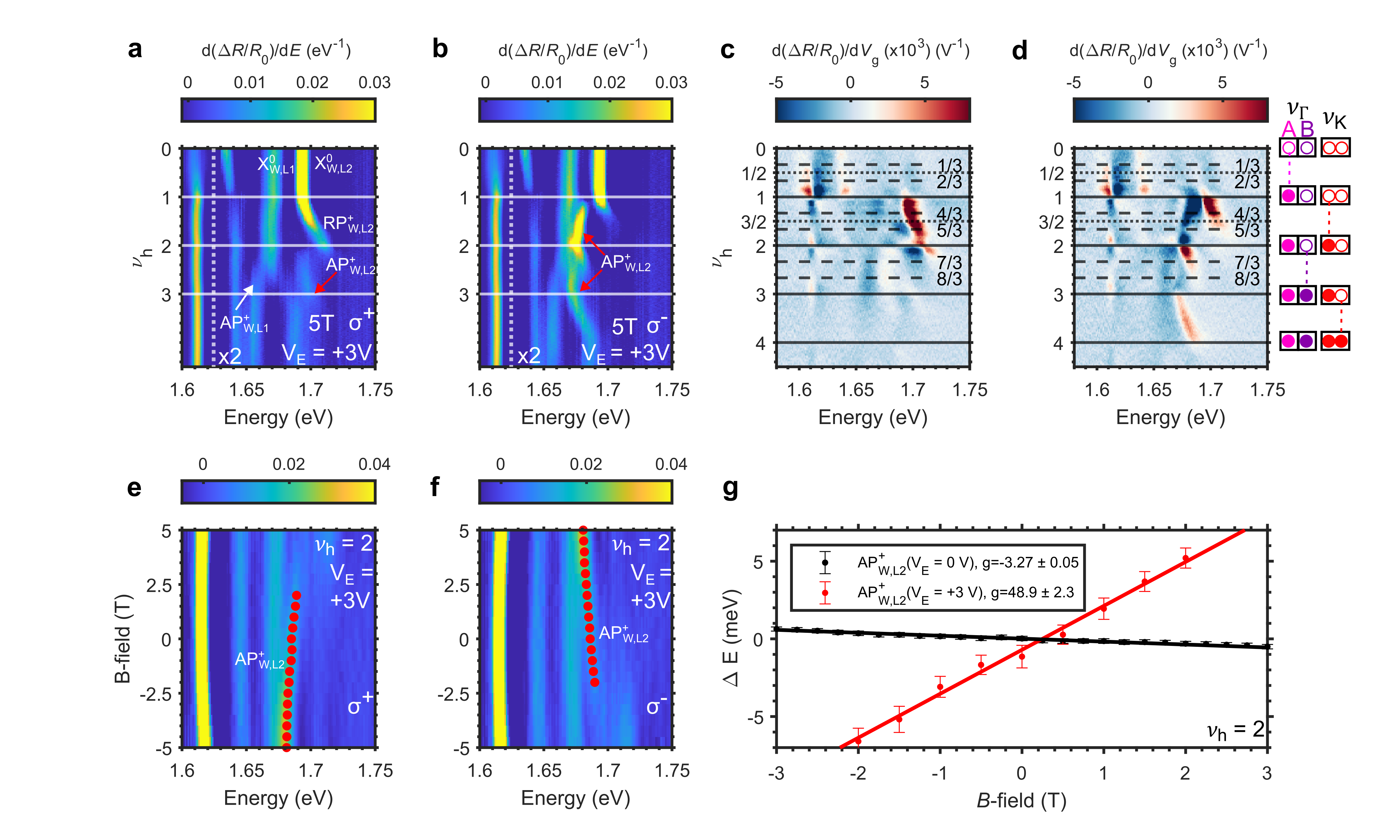}
    	\end{center}
        \caption{\textbf{Inducing spin and valley polarisation in K orbitals with $\vec{E}$}. \textbf{a,b}, $\nu_h$ dependence of the $\sigma^+$ (a) and $\sigma^+$ (b) resolved d($\Delta R/R_0$)/d$E$ at $V_\mathrm{E} = +3$~V under application of a 5~T magnetic field.  \textbf{c,d}, Density plot of d($\Delta R/R_0$)/d$V_\mathrm{g}$ in panel (a) and (b), respectively, which highlight the formation of correlated states, as identified by the dashed and solid lines. The orbital at the Fermi level is depicted in the schematic to the right. \textbf{e,f} $\sigma^+$ (e) and $\sigma^-$ (f) resolved $B$-field sweep of d($\Delta R/R_0$)/d$E$ at $V_\mathrm{E} = +3$~V and $\nu_h =2$. The red dots represent the magnetic-field-dependent estimated energies of the AP$_\mathrm{W,L2}^+$. \textbf{g} $B$-field-dependent Zeeman splitting, $\Delta E$, of AP$_W^+$ at representative hole $\nu_h = 2$ values for $V_\mathrm{E} = 0$~V (black dots) and $V_\mathrm{E} = +3$~V (red dots). For $V_\mathrm{E} = +3$~V, $\Delta E$ is extracted from the estimated positions shown in panels (e,f). The $g$-factor of AP$_\mathrm{W,L2}^+$ is extracted from linear fits to $\Delta E$ (solid lines).} \label{fig:3}
\end{figure*}

We now investigate the field tunability of the bands. By probing the magneto-optical properties of the WSe$_2$ exciton-polarons, we disentangle the different spin and valley properties of the holes doped into the $\Gamma$ and $\pm$K moir{\'e} bands as a function of $\vec{E}$. Figure~\ref{fig:3}a,b shows polarisation ($\sigma^+$- and $\sigma^-$, respectively)-resolved d($\Delta R/R_0)$/d$E$ sweeps as a function of $\nu_h$ with B = 5~T and $V_\mathrm{E}= +3$~V.  As we increase $\nu_h$ in the $\sigma^-$-resolved spectrum (Fig.~\ref{fig:3}b), a new resonance (labeled AP$^+_\mathrm{W,L2}$, indicated by the red arrow) appears at $\nu_h = 1$, $\sim 14$~meV lower energy than X$^0_\mathrm{W,L2}$. This resonance is notably absent for the $\sigma^+$ spectrum (Fig.~\ref{fig:3}a), where instead the X$^0_\mathrm{W,L2}$ evolves into the repulsive-polaron branch RP$^+_\mathrm{W,L2}$ which blueshifts with increasing doping. The difference between the $\sigma^+$ and  $\sigma^-$-resolved sweeps indicates a large degree of spin-polarisation for the carriers that form the exciton-polarons in L2. We therefore unambiguously identify AP$^+_\mathrm{W,L2}$ as the inter-valley attractive polaron which is formed from the direct $\pm$K exciton in L2 dressed with holes in the opposite K valley in L2 \cite{liu2020landau,wang2020observation}. On application of the positive magnetic field, the valence band edge in L2 at $-$K is shifted to higher energy relative to $+$K \cite{srivastava2015valley,liu2020landau,wang2020observation}. Exchange interactions favour single valley occupancy of carriers: all the holes are doped into the $-$K rather than $+$K valley, leading to the clear spin-polarisation \cite{back2017giant}. When we probe the exciton in $+$K using $\sigma^-$ light, there is a large population of holes available to form the attractive polaron. In contrast, there is no population of holes at $+$K available to form the attractive polaron when we probe the $-$K exciton using $\sigma^+$. This observation strongly indicates that at $V_\mathrm{E}= +3$~V, the L2 K-derived band is tuned upwards in energy relative to the layer hybridised $\Gamma_\mathrm{B}$ band. Previously, for $V_\mathrm{E} = 0~$V and holes loaded only at $\Gamma$, there is a complete lack of hole spin-polarisation, regardless of the fractional filling (see Fig.~\ref{fig:2}a,b).

Next, we investigate the interaction effects on the magneto-optics for the correlated states in the $\Gamma$ and K orbitals. Figure \ref{fig:3}e and \ref{fig:3}f show the $\sigma^-$- and $\sigma^+$-helicity-resolved evolution of the d$(\Delta R/R_0)$/d$E$ spectrum for applied $B$-fields between $-5$ T and 5 T at $\nu_h = 2$ and $V_\mathrm{E} =+3$~V. The $B$-field-dependent estimated energy of AP$_\mathrm{W,L2}^+$ extracted from fits are overlayed (red dots). Figure \ref{fig:3}g shows the Zeeman splitting, $\Delta E$, of AP$_\mathrm{W,L2}^+$ versus $B$-field at $\nu_h = 2$, for both $V_\mathrm{E} = 0~$V (black) and $V_\mathrm{E} = +3$~V (red), where $\Delta E=E^{\sigma^+}-E^{\sigma^-}$, with $E^{\sigma^{\pm}}$ the energy of the transition with $\sigma^{\pm}$ polarisation. The evolution of $\Delta E$ with $B$-field can be associated with an effective exciton $g$-factor, $g$, using  $\Delta E(B) = g\mu_0B$, where $\mu_0$ is the Bohr magneton. For $V_\mathrm{E} = 0~$V, when AP$^+_\mathrm{W,L2}$ is formed by an exciton dressed by holes at $\Gamma$, the $g$-factor is close to $-4$, which is expected according to spin and valley considerations in the absence of interaction effects \cite{forste2020exciton}. In contrast, for $V_\mathrm{E} = +3$~V, AP$^+_\mathrm{W,L2}$ exhibits a large positive $g$-factor, $\sim49$, partly due to the aforementioned interaction effects that arise from hole doping into the $\pm$K band and the resulting favoured single valley occupancy (see Suppl. Fig. S3 and Suppl. Tab. S2) \cite{back2017giant}. The measured $g$-factor is then further enhanced by spin-coupling between the moir{\'e} localised holes \cite{tang2020simulation,campbell2022strongly}.  

The tuning of the valence bands in the system is further demonstrated by the suppression of the formation of the attractive polaron for the L1 WSe$_2$ exciton, AP$^+_\mathrm{L1}$, up to $\nu_h > 2$: holes are doped into the highly localised K-L2 band and therefore have little effect on the L1 WSe$_2$ exciton. Finally, we note that the binding energy, $E_\mathrm{B}$, for the hole-dressed attractive polaron formed by the L2-localised exciton for $V_\mathrm{E} = +3$~V ($E_\mathrm{B} \sim$ 14~meV) is much larger than the binding energy for $V_\mathrm{E} = 0$~V ($E_\mathrm{B} \sim$ 6~meV). This result is consistent with a larger wavefunction overlap between the exciton at $\pm$K dressed by holes at $\mp$K compared to holes at $\Gamma$. 

\section{Correlated states: $\Gamma$ versus K orbitals}

We now investigate the correlated states formed within the $\Gamma$ and K orbitals. We expect the same dependence of $V_\mathrm{g}$ versus $\nu_h$ for each orbital. Figure~\ref{fig:3}c,d, shows the $\nu_h$ dependence of d($\Delta R/R_0$)/d$V_\mathrm{g}$ for the same data in panels (a) and (b), respectively. As expected, we observe the largest changes in the d($\Delta R/R_0$)/d$V_\mathrm{g}$ signal at integer $\nu_h$ and further smaller changes at intermediate $\nu_h$, as previously observed for $V_\mathrm{E} = 0$~V (see Fig.~\ref{fig:2}c,d). However, we observe that new correlated states emerge for $V_\mathrm{E} = +3~$V compared to $V_\mathrm{E} = +0~$V. A new state at $\nu_h = 4$ clearly appears, as well as further states at $\nu_h = 7/3$ and $8/3$. 

To understand the origin of the additional correlated insulating states, we again use the polarisation-resolved doping-dependent reflection contrast measurements (see Fig.~\ref{fig:2}a,b and Fig.~\ref{fig:3}a,b), and assign the correlated states to filling of either the $\Gamma$ or K orbitals, as indicated by the schematics at the far right of the figures. For $\nu_h<1$, we observe that attractive polarons do not form for any $V_\mathrm{E}$ because holes are always first doped into the $\Gamma_\mathrm{A}$ orbital. We then observe that AP$^+_\mathrm{W,L2}$ and RP$^+_\mathrm{W,L2}$ form at $\nu_h>1$ with a large spin-polarisation demonstrating doping into the K orbital. Hence, for $V_\mathrm{E} = +3$~V at $\nu_h = 2$, a correlated state arises in which the lower Hubbard bands are filled for both the $\Gamma_\mathrm{A}$ and K orbitals. In contrast, for $V_\mathrm{E} = 0$~V and $1<\nu_h<2$, the $\Gamma_\mathrm{B}$ orbital is filled, as demonstrated by the lack of spin-polarisation of the carriers that form AP$^+_\mathrm{W,L1}$ and AP$^+_\mathrm{W,L2}$ (see Fig.~\ref{fig:2}a,b).

Continuing with the $V_\mathrm{E} = +3$~V scenario, after $\nu_h = 2$ the non-spin-polarised AP$^+_\mathrm{W,L1}$ forms because holes are now doped into the $\Gamma_\mathrm{B}$ orbital. Hence, at $\nu_h = 3$, a correlated state arises in which each orbital ($\Gamma_\mathrm{A}$, $\Gamma_\mathrm{B}$, and K) is filled with one hole. Finally for $\nu_h > 3$, AP$^+_\mathrm{W,L2}$ continually blueshifts with increasing doping, indicating Pauli blocking effects arising from doping into the K orbital, whereas AP$^+_\mathrm{W,L1}$ remains unaffected. At $\nu_h = 4$, we therefore have a correlated state with $\nu_\Gamma=2$ and $\nu_\mathrm{K}=2$. These assignments are consistent with the emergence of new states at $\nu_h = 7/3$ and $8/3$ for $V_\mathrm{E} +3$~V. In comparison, for $V_\mathrm{E} = 0~V$, the Wigner crystals driven by long-range interactions are not observed at these fractional fillings of the moir{\'e} lattice in the next $\Gamma$ moir{\'e} orbital, either due to more disorder or weaker correlations (increased bandwidth). The emergence of new correlated states at $\nu_h = 7/3, 8/3$ and $4$ is consistent with metal-to-insulator transitions occurring at these fractional fillings, driven by $V_\mathrm{E}$-dependent tuning of the relative $\Gamma-$K band energies. 

\begin{figure*}[htp]
    	\begin{center}
    		\includegraphics[scale= 0.45]{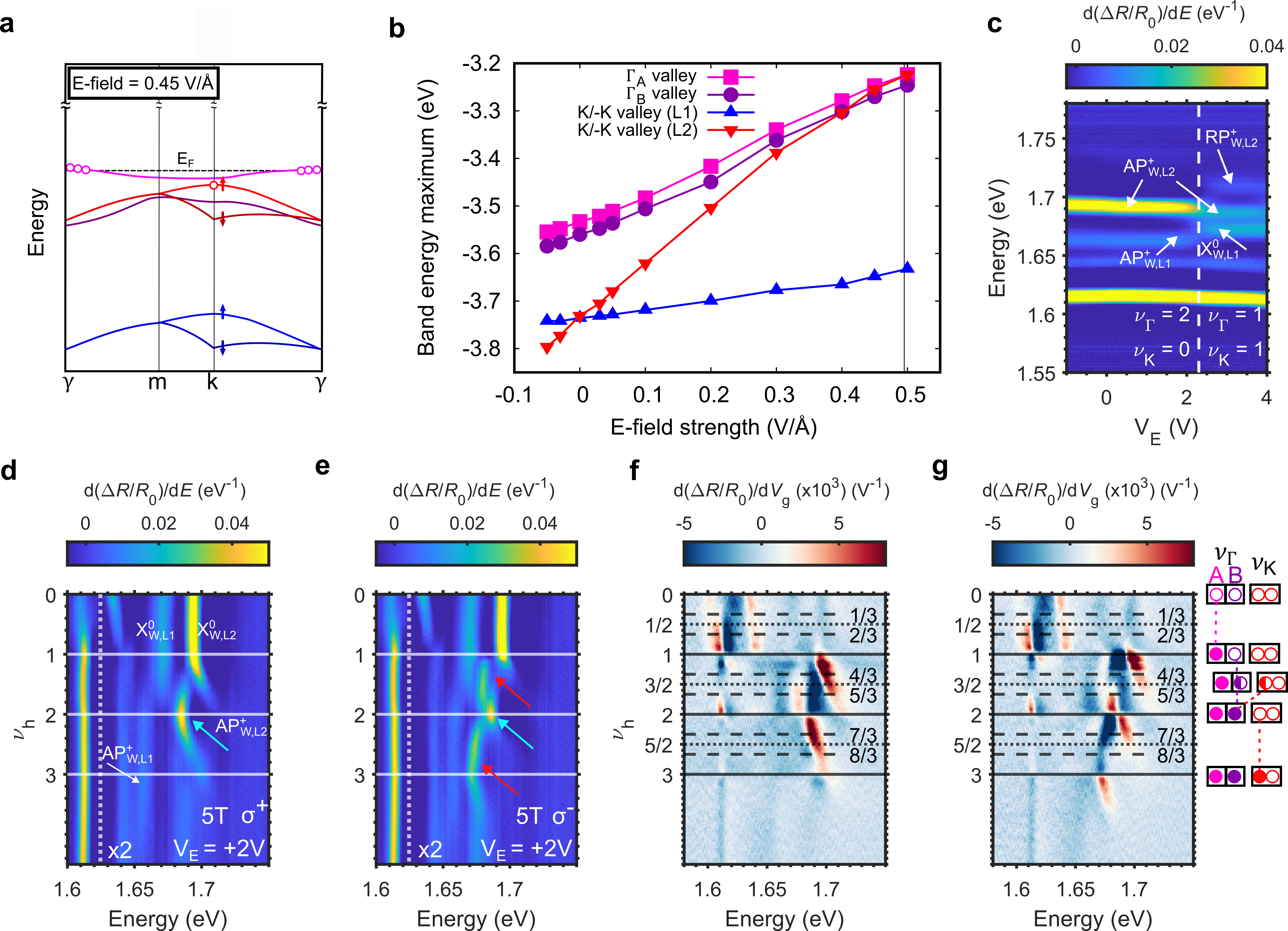}
    	\end{center}
        \caption{\textbf{Interplay of K and $\Gamma$ correlated states.} \textbf{a} Band structure schematic for the moir{\'e} Brillouin zone based on DFT calculations of the heterotrilayer under an applied positive vertical electric field. The K-derived band is shifted to higher energy relative to the $\Gamma$-derived bands compared to $V_\mathrm{E}$ = 0. \textbf{b} Electric field dependence of the valence band maximum for the $\Gamma$ and K-derived moir{\'e} bands, calculated from DFT. \textbf{c} Density plot of the $V_\mathrm{E}$ dependence of the derivative of the d($\Delta R/R_0$)/d$E$ when the hole filling is fixed at $\nu_h = 2$. The white dashed line indicates an abrupt transfer of holes in the system from the $\Gamma$-derived moir{\'e} orbital to the K-derived orbital. \textbf{d,e} $\nu_h$ dependence of the $\sigma^+$ (d) and $\sigma^-$ (e) resolved d($\Delta R/R_0$)/d$E$ at $V_\mathrm{E} = +2$~V under application of a 5~T magnetic field. \textbf{f,g} Density plot of d($\Delta R/R_0$)/d$V_\mathrm{g}$ for the same data in panels (d) and (e), respectively. On all plots, the solid lines indicate integer hole fractional fillings of the moir{\'e} superlattice, while the dashed and dotted lines in (f,g) show intermediate fractional fillings. The schematics on the far right indicate the carriers per moir{\'e} unit cell in each of the $\Gamma$-derived or K-derived orbitals, where a filled circle indicates integer filling. }
        \label{fig:4}
\end{figure*}

\section{The interplay of correlated states between degenerate orbitals}

To better understand the different $V_\mathrm{E}$-dependent shifts in the K and $\Gamma_\mathrm{B}$ orbitals, we perform DFT calculations for the HTL in the presence of a vertical electric field. Figure~\ref{fig:4}a shows a simplified band structure schematic in the moir{\'e} Brillouin zone for a positive vertical electric field (see Suppl. Fig. S10) for calculated full band structure). In agreement with the experimental results, we observe that the $\pm$K-derived bands which are localized on L2 are pushed up in energy (i.e. reduced hole energy) relative to the $\Gamma_\mathrm{B}$ bands (see Fig.~\ref{fig:1}e for a comparison). Fig.~\ref{fig:4}b shows the band-edge energy of the $\Gamma_\mathrm{A}$, $\Gamma_\mathrm{B}$, K-L1, and K-L2 bands as function of the applied electric field strength. Due to the high localisation of its wavefunction on L2 WSe$_2$ (see Suppl. Fig. S10), the K-L2  orbital exhibits the largest $V_\mathrm{E}$ dependent shift in energy. In contrast, the $\Gamma_\mathrm{A}$ and $\Gamma_\mathrm{B}$ bands exhibit smaller $V_\mathrm{E}$ dependent energy shifts due to the delocalisation across both WSe$_2$ layers\cite{movva2018tunable}. 

To probe the interplay between correlated states in $\Gamma$ and K orbitals in more detail, we measure the $V_\mathrm{E}$ dependence of the d($\Delta R/R_0)/$d$E$ signal for a fixed hole filling of $\nu_h = 2$ (see Fig.~\ref{fig:4}c). For $V_\mathrm{E}=0~$V, we observe the two WSe$_2$ attractive polaron resonances: intralayer excitons localised in L1 and L2 of the bilayer dressed by delocalised holes residing in the highest $\Gamma$ moir{\'e} miniband (AP$^+_\mathrm{W,L1}$ and AP$^+_\mathrm{W,L2}$, respectively). However, as $V_\mathrm{E}$ is increased, these features are abruptly replaced by 3 new resonances at $V_\mathrm{E}{\sim}2.3~$V, indicated by the vertical white dashed line. The full width half maximum of the change in the d($\Delta R/R_0)/$d$E$ signal is calculated to be $\Delta V_\mathrm{E} = 0.53$~V (see Suppl. Fig. S2). To identify the origin of the abrupt change in the excitonic resonances, we measure the $\sigma^+$- and $\sigma^-$- resolved $\nu_h$ dependence of the d($\Delta R/R_0)$/d$E$ signal, respectively, with $V_\mathrm{E} = +2~$V (near the abrupt transition in Fig.~\ref{fig:4}c) and $B=$ 5~T (see Fig.~\ref{fig:4}d,e). Figure~\ref{fig:4}f,g, shows the $\nu_h$ dependence of d($\Delta R/R_0$)/d$V_\mathrm{g}$ for the data in panels (d) and (e), respectively. Here, the large change at $\nu_h = 2$ in d($\Delta R/R_0$)/d$V_\mathrm{g}$ indicates strongly correlated state formation.  Identical to the case for $V_\mathrm{E} = +3~$V, in the $\sigma^-$ spectrum (Fig.~\ref{fig:4}d) we observe the emergence of a new highly spin-polarised resonance AP$^+_\mathrm{W,L2}$ (red arrow), indicating holes are initially doped into the L2 K moir{\'e} band for $\nu_h>1$. However, for $\sigma^-$, around $\nu_h = 2$ the oscillator strength is transferred to a higher energy feature with a smaller binding energy, which also appears for the $\sigma^+$-resolved data (indicated by the blue arrows). This feature lacks a strong degree of spin-polarisation, suggesting that holes are transferred from the K to the $\Gamma_\mathrm{B}$ orbital at $\nu_h=2$. The hole transfer is further demonstrated by the formation of the attractive polaron for the L1 WSe$_2$ exciton around $\nu_h = 2$, at the same hole filling where the oscillator strength transfer occurs. For $\nu_h > 2$, holes start to be doped into the K orbital again, and the spin-polarised features re-emerge. Therefore, we assign the abrupt change in the the absorption spectrum at $V_\mathrm{E} = +2.3~$V in Fig.~\ref{fig:4}c to a transfer of holes from the $\Gamma_\mathrm{B}$ to the K moir{\'e} orbital. 

Rather than being continuous, the observed change occurs over a small range of $V_\mathrm{E}$ ($\Delta V_\mathrm{E} \sim 0.5~V$).  If holes were continuously transferred between the orbitals there would be a gradual change in oscillator strength \cite{shimazaki2020strongly}. Instead, the abrupt change indicates that the system favours correlated state formation, and the different $V_\mathrm{E}$-dependent scenarios for $\nu_h = 2$ provide an archetypal illustration. For small $V_\mathrm{E}$, a correlated state with $\nu_h = 1$ in each $\Gamma_\mathrm{A}$ and $\Gamma_\mathrm{B}$ orbital arises. For large $V_\mathrm{E}$, a correlated state with $\nu_h = 1$ in the $\Gamma_\mathrm{A}$ orbital and $\nu_h = 1$ in the K orbital is formed. However, when the $\Gamma_\mathrm{B}$ and K orbitals are nearly degenerate at the Fermi level, the holes preferentially transfer from the K orbital to the $\Gamma_\mathrm{B}$ orbital, likely due to a larger Coulomb gap for the filled $\Gamma_\mathrm{B}$ orbital which minimises the energy of the correlated state. A summary of the orbital filling near degeneracy is depicted in the schematic to the right of Fig.~\ref{fig:4}g.

\section{Discussion}

In the experiment, we observe an interplay between correlated states in three orbitals which can be tuned with $V_\mathrm{E}$. Here we develop a theoretical model to provide a more detailed picture of the charge ordering and interactions of these correlated states. Charges can be distributed (and redistributed) between the $\Gamma_\mathrm{A}$, $\Gamma_\mathrm{B}$, and K orbitals as hole filling is increased. It is likely that complex correlated states with specific charge ordering in real space arise to minimise electrostatic repulsion for a given $\nu_h$, similar to that observed in bilayer Hubbard model physics \cite{xu2022tunable}. 

\begin{figure*}[htp]
    	\begin{center}
    		\includegraphics[scale= 0.5]{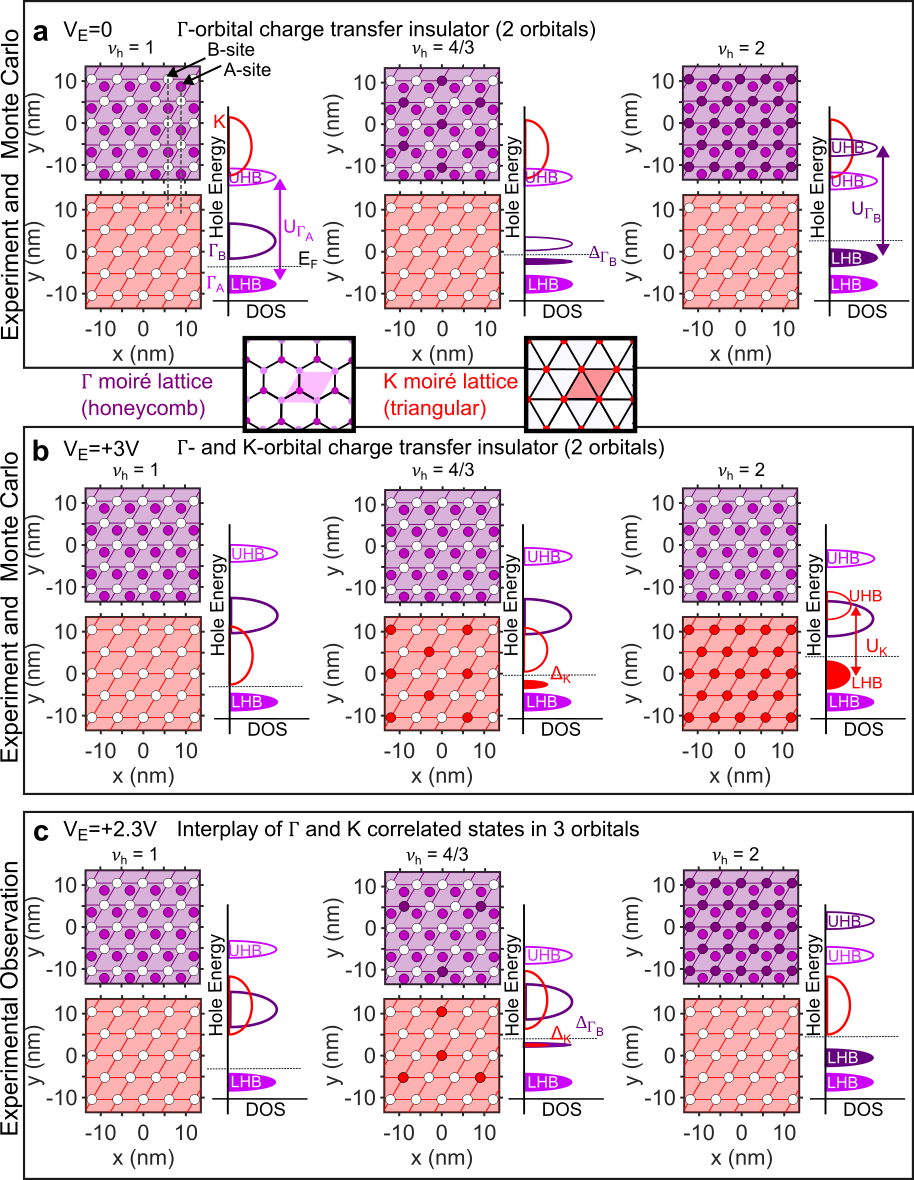}
    	\end{center}
        \caption{\textbf{Real-space charge distributions of correlated states in the $\Gamma_\mathrm{A}$, $\Gamma_\mathrm{B}$, and K orbitals at selected values of $\nu_h$ and $V_\mathrm{E}$.} \textbf{a} Monte Carlo simulated charge ordered states that arise in the $\Gamma$ (pink and purple) and K (red) moir{\'e} lattices for $V_\mathrm{E} = 0$, for which the energy of the K-derived band is below both $\Gamma-$ derived bands ($\epsilon_\mathrm{K}$ < $\epsilon_\mathrm{B}$). The $\Gamma_\mathrm{B}$ sites are spatially aligned with the K sites in the $x-y-$ plane. Holes are first loaded into the $\Gamma_\mathrm{A}$ sites up to $\nu_h = 1$ and into the $\Gamma_\mathrm{B}$ sites for $\nu_h > 1$. Schematics to the right illustrate the hole energy versus density of states (DOS) for each charge ordered state depicted. Lower and upper Hubbard bands (LHB and UHB) form for $\nu_h = 1$, while a charge gap $\Delta$ opens for correlated states that form for non-integer filling. \textbf{b} Monte Carlo simulated charge ordered states and corresponding DOS schematics that arise for $V_\mathrm{E} = +3$~V, for which the energy of the K orbital lies between the $\Gamma$ orbitals ($\epsilon_\mathrm{A}$ > $\epsilon_\mathrm{K}$ > $\epsilon_\mathrm{B}$). Holes are first loaded into the $\Gamma_\mathrm{A}$ sites up to $\nu_h = 1$, and into K for $\nu_h > 1$. \textbf{c} Experimentally observed charge ordered states and corresponding DOS schematics for $V_\mathrm{E} = +2.3$~V for which the K orbital is energetically degenerate with the $\Gamma_B$ orbital.  Holes are first loaded into the $\Gamma_\mathrm{A}$ sites up to $\nu_h = 1$, and into both K and $\Gamma_\mathrm{B}$ for $\nu_h > 1$. At $\nu_h = 2$ holes are abruptly transferred from K to $\Gamma_\mathrm{B}$ to form the correlated state with $\nu_{\Gamma \mathrm{A}} = 1$ and $\nu_{\Gamma \mathrm{B}} = 1$.}
        \label{fig:5}
\end{figure*}

We analyze a three-orbital $t$-$U$-$\mathcal{V}$ Hamiltonian in the $t/U \rightarrow 0$ limit. In this limit, each moir\'e site can be occupied by at most one hole and no hopping between moir\'e sites is allowed. To find the configuration of holes $n_i$ (where $i$ labels the various moir\'e sites corresponding to L2 K, $\Gamma_\mathrm{A}$, and $\Gamma_\mathrm{B}$ orbitals) which minimizes the total energy, we use a Monte Carlo simulated annealing approach. The total energy is given by 
\begin{eqnarray}
    H &=& \varepsilon_\mathrm{A}\sum_{\substack{i}\in \Gamma_\mathrm{A}} n_{i}+
    \varepsilon_\mathrm{B}\sum_{\substack{i}\in \Gamma_\mathrm{B}} n_{i}+
    \varepsilon_\mathrm{K}(V_\mathrm{E})\sum_{\substack{i}\in \mathrm{K}} n_{i} \nonumber\\ & & + \frac{1}{2}\sum_{\substack{i,j} \in \{\Gamma_\mathrm{A},\Gamma_\mathrm{B},\mathrm{K}\}} \mathcal{V}(|\mathbf{r}_i-\mathbf{r}_j|) n_{i}n_{j},\label{eq:Hamiltonian}
\end{eqnarray}
where $\varepsilon_\mathrm{A}$,  $\varepsilon_\mathrm{B}$ and $\varepsilon_\mathrm{K}(V_\mathrm{E})$ denote the onsite energies of the $\Gamma_\mathrm{A}$, $\Gamma_\mathrm{B}$, and L2 K orbitals, respectively. Note that the onsite energy of the L2 K orbital (relative to that of the $\Gamma$ orbitals)  depends on the applied electrostatic potential $V_\mathrm{E}$. From our DFT calculations, we estimate $\varepsilon_\mathrm{A} - \varepsilon_\mathrm{B} \approx 20$~meV, independent of $V_\mathrm{E}$.
Also, \mbox{$\mathcal{V}(|\mathbf{r}_i-\mathbf{r}_j|) = \frac{\exp{(-|\mathbf{r}_i-\mathbf{r}_j|/L_0)}}{\epsilon|\mathbf{r}_i-\mathbf{r}_j|}$} denotes the screened Yukawa interaction between holes located at positions $\mathbf{r}_i$ and $\mathbf{r}_j$, where $\epsilon$ is the dielectric constant which we set to 4.5 and $L_0$ is the characteristic screening length which we set to $2$~nm. As the K-derived states are highly localized on the WSe$_2$ L2 layer, the corresponding hole positions are assumed to lie in the plane of the layer. In contrast, our DFT calculations show that the $\Gamma$-derived states are delocalized across both WSe$_2$ layers and we therefore assume that the corresponding hole positions lie at the midpoint between the layers.

First, we model the trilayer without an electric field, i.e. $V_E=0$. In this case, $\varepsilon_\mathrm{K}$ is much larger than both $\varepsilon_\mathrm{A}$ and $\varepsilon_\mathrm{B}$. For $\nu_h = 1$, the holes occupy the $\Gamma_\mathrm{A}$ orbitals, forming a triangular lattice as shown in Fig.~\ref{fig:5}a. The electronic structure of this correlated state is characterized by an occupied lower hole Hubbard band which is separated from the upper hole Hubbard band by the onsite Coulomb energy $U_{\Gamma_A}$, see Fig.~\ref{fig:5}a. Further hole doping results in occupancy of the $\Gamma_\mathrm{B}$ sites, and to the formation of charge-ordered $\Gamma_\mathrm{B}$-states, consistent with the experimental findings of the $\Gamma$-derived Wigner crystal at $\nu_h=4/3$ (shown in Fig.~\ref{fig:2}c,d). At $\nu_h=2$, all $\Gamma_\mathrm{A}$ and $\Gamma_\mathrm{B}$ sites are occupied while the L2 K orbital remains empty. 

Next, we consider the case of a large electric field, $V_E=+3$~V. In this case, $\varepsilon_\mathrm{B} > \varepsilon_\mathrm{K} > \varepsilon_\mathrm{A}$ and for $\nu_h=1$ the holes still occupy the $\Gamma_\mathrm{A}$ sites. Further hole doping, however, results in the L2 K orbital becoming occupied and forming charge-ordered states, see Fig.~\ref{fig:5}b, consistent with the experimental picture of $B$-field-dependent correlated states at $\nu_h=4/3$. At $\nu_h=2$, the lower Hubbard bands of the $\Gamma_\mathrm{A}$ and L2 K orbitals are fully occupied by holes while the upper Hubbard bands are empty.  




Finally, we consider the critical electric field, $V_E=+2.3$~V. At this field, the $\varepsilon_\mathrm{B} \approx \varepsilon_\mathrm{K} > \varepsilon_\mathrm{A}$ and we expect strong quantum interaction effects to occur which are not captured by our electrostatic model. For $\nu_h=1$, the holes still occupy the $\Gamma_\mathrm{A}$ sites. As a result of the degeneracy of the $\Gamma_\mathrm{B}$ and the L2 K-orbitals, we hypothesize that additional holes ($1<\nu_h$ <2) occupy both orbitals to minimize the overall Coulomb repulsion, see Fig.~\ref{fig:5}c. When the doping reaches $\nu_h=2$, we experimentally observe that all holes reside in $\Gamma$-derived states. This can be explained by the formation of a correlated state in which the lower Hubbard bands of both $\Gamma_\mathrm{A}$ and $\Gamma_\mathrm{B}$ orbitals are fully occupied. As the bandwidth of the $\Gamma$-derived bands is smaller than that of the K-derived bands (according to our DFT calculations), compared to a state in which the L2 K orbital is fully occupied it is energetically favorable for the $\Gamma$ orbitals to be fully occupied. Thus, the holes spontaneously redistribute from the K to the $\Gamma_\mathrm{B}$ orbital. This observation provides motivation for future theoretical and experimental investigations to fully understand the many-body physics of this strongly correlated multi-orbital system.

\section{Conclusions}

In conclusion, we use exciton-polaron spectroscopy of a moir{\'e} HTL device to reveal strongly correlated hole states in $\Gamma$ and K orbitals that are best described by an extended $t$-$U$-$\mathcal{V}$ three-orbital Hubbard model. Notably, we find that the honeycomb $\Gamma$-band gives rise to a charge-transfer insulator which also hosts generalized Wigner crystal states at several intermediate fractional fillings.  With application of $\vec{E}$, we change the ordering of the $\Gamma_\mathrm{B}$- and K-derived bands and observe a sudden transfer of the holes from the correlated states formed in the $\Gamma_\mathrm{B}$ to the K orbital, where new strongly correlated states emerge. Further, in the critical condition when when $\Gamma_\mathrm{B}$ and K are degenerate at the Fermi level, we observe stable correlated states simultaneously in each orbital at non-integer fractional fillings (between $1<\nu_h<2$). However, upon approaching $\nu_h$ = 2, the holes abruptly redistribute from the K to the $\Gamma_\mathrm{B}$ orbital, likely due to a larger Coulomb gap for the filled $\Gamma_\mathrm{B}$ orbital which minimises the energy of the correlated state. 

Our work establishes the HTL as a platform for investigation of both honeycomb and triangular Hubbard systems, as well as their interplay \cite{kennes2021moire}. Exploring the magnetic phase diagrams of the charge transfer insulating states \cite{zhang2020moire,slagle2020charge} and  honeycomb lattice Wigner crystals \cite{kaushal2022magnetic,xian2021realization} are exciting future avenues to pursue. Further, the high tunability and large number of degrees of freedom available in this system could enable the exploration of Kondo interactions when $\nu_h = 1$ in the lowest Hubbard band while a higher energy, more dispersive orbital is partially filled \cite{dalal2021orbitally,kumar2022gate}.

\section{Methods}

\subsection{Exciton-Polaron Spectroscopy of Correlated States With Layer and Valley Specificity}

To precisely probe the nature of the correlated states, we develop an optical spectroscopy technique that yields unambiguous layer and valley specific experimental signatures. We probe momentum-direct electrons and holes at $\pm$K valleys that form tightly-bound intralayer excitons which are extremely sensitive to their environment. The exciton binding energy depends on its dielectric environment \cite{raja2017coulomb}, leading to layer specific exciton energies \cite{chen2022excitonic}, while excitons dressed by a Fermi sea form attractive (AP) and repulsive (RP) exciton-polarons \cite{sidler2017fermi,roch2019spin,efimkin2017many,fey2020theory} which exhibit distinct behaviours depending on the valley hosting the Fermi sea. For doping at $\pm$K, dominant phase space filling effects cause an overall blue-shift in the AP \cite{back2017giant,wang2017probing} while doping in the $\Gamma$ valley leads to a continuous red-shift due to bandgap renormalisation \cite{gao2016dynamical,yao2017optically}. Contrasting properties in an applied magnetic field also manifest for excitons dressed by carriers in the two valleys: unlike $\Gamma$-valley holes, K-valley holes are highly spin-polarised and exhibit strong magnetic interactions with excitons at $\pm$K \cite{back2017giant}.  Finally, the formation of strongly correlated states leads to abrupt changes in the oscillator strength, energy, and linewidth of the exciton polarons, regardless of which valley the Fermi-sea occupies \cite{campbell2022strongly,tang2020simulation,shimazaki2020strongly}. Altogether, exciton-polarons provide probes of strongly correlated electronic states with layer and valley specificity.

\subsection{Device Details}

The heterostructure is fully encapsulated in hexagonal boron nitride (hBN). There are few layer graphene contacts to both the top and bottom hBN, as well as the heterostructure which allows the carrier concentration and vertical electric field to be tuned independently (see Fig.~\ref{fig:1}b). To change the carrier concentration in the heterostructure, without changing the vertical electric field, we sweep both $V_T$ and $V_\mathrm{B}$ with the same values, i.e.  $V_T = V_\mathrm{B} = \Delta V_\mathrm{g}$. To change the vertical electric field without changing the carrier concentration, $V_T = -V_\mathrm{B} = V_\mathrm{E}$. The applied vertical electric field can then be calculated using $E = \frac{(V_T - V_\mathrm{B})}{(d_1+d_2)} \frac{\epsilon_{hBN}}{\epsilon_{TMD}}$. For intermediate regimes, when we want to change both the applied vertical electric field \emph{and} the carrier concentration, we apply asymmetric gating, i.e. $V_T \neq V_\mathrm{B}$ \cite{jauregui2019electrical}. 

The carrier concentration $n$ in the heterostructure can be calculated using the parallel plate capacitance model, $n= \frac{\epsilon\epsilon_0 \Delta V_\mathrm{g}}{d_1}+ \frac{\epsilon\epsilon_0 \Delta V_\mathrm{g}}{d_2}$ where $\epsilon$ is the permittivity of hBN, $\Delta V_\mathrm{g}$ is the voltage offset between both the top and bottom gates and the heterostructure gate, and $d_1$ ($d_2$) is the thickness of the top (bottom) hBN layer measured to be $17.4\pm0.2$ nm ($18.2\pm0.3$ nm) using nulling ellipsometry. For a small angular difference between two stacked layers the moir{\'e} periodicity can be estimated using $\lambda_M = \frac{a_{Se}}{\sqrt{\delta^2+\theta^2}}$  where $a_{Se}$ is the lattice constant of WSe$_2$, $\delta$ is the fractional lattice mismatch between the two layers and $\theta$ is the twist angle in radians. For a triangular moir{\'e} pattern (or honeycomb with inequivalent A and B sites, where only one site is doped), the number of carriers required for one hole per site is given by $n_0=\frac{2}{\sqrt{3}\lambda_{M}^{2}}$. Using the lattice constants of 0.3280~nm and 0.3288~nm for WSe$_2$ and MoSe$_2$ respectively \cite{brixner1962preparation} and a permitivitty of 3.8 for the hBN \cite{laturia2018dielectric}, and $\Delta V_\mathrm{g} = 1.00$~V for $\nu_h$ $= 1$ as determined in the main text, we calculate the twist angle in the main measurement location to be 2.7$^\circ$. We determine the interface in the HTL is R-stacked, as a result of the fabrication process. Single flakes of MoSe$_2$ and WSe$_2$ with monolayer and natural 2H bilayer regions are first exfoliated with the terraced layers (flat surface) facing up (on the substrate). The WSe$_2$ flake (terraced side up) is then transferred onto the MoSe$_2$ on its substrate, keeping the interface between the two layers free from contaminants. The HBL region is H-stacked, as demonstrated by magneto-optical studies on the same sample \cite{baek2020highly,brotons2021moir} while  in the HTL region the WSe$_2$ interface layer L1 is R-stacked relative to the MoSe$_2$ monolayer. The upper WSe$_2$ layer L2 is therefore H-stacked relative to the monolayer MoSe$_2$. We note that second harmonic generation measurements cannot be used to determine R- versus H-stacking of the hetero-interface of the HTL.

\subsection{Optical Spectroscopy Measurements}

We focus a broadband white light source from a power-stabilised halogen lamp onto the sample, and measure the reflected signal in a liquid nitrogen cooled CCD spectrometer. The sample is mounted in a closed-cycle cryostat and the sample temperature for all experiments in the main text is 4~K. We define reflection contrast as $\Delta R/R_0$, where $\Delta R = (R_S-R_0)/R_0$. Here $R_S$ is the reflected signal from the heterostructure, and $R_0$ is the reflected signal from a nearby heterostructure region without the TMD layers.

\subsection{DFT calculations}
The moir\'e unit cell for the free-standing 2$H$ WSe$_2$/MoSe$_2$ heterotrilayer with a twist angle of $3.1^\circ$ contains 2979 atoms. To calculate the band structure, we first relax the atomic positions using classical force fields as implemented in the LAMMPs code\cite{LAMMPS} (see Suppl. Sec. S2 for details on the heterotrilayer geometry and atomic relaxations). DFT calculations were carried out on the relaxed structure using the linear-scaling DFT code SIESTA \cite{Jose_M_Soler_2002}. We employ the LDA exchange-correlation functional \cite{LDA}, with a double-zeta plus polarization (DZP) pseudoatomic orbitals basis set, a mesh cutoff of 300~Ry and a vacuum region between periodic images of  15~\AA. Spin-orbit coupling was included via the ``on-site'' approximation\cite{Fernandez_Seivane_2006}. Electron-ion interactions are described by fully relativistic Troulliers-Martins pseudopotentials \cite{TroulliersMartins1991}. For the calculations with a non-zero vertical electric field, the same atomic positions as in the zero-field case are used and we employ a slab dipole correction, in which the electric field used to compensate the dipole of the system is computed self-consistently in every iteration \cite{PAPIOR20178}. DFT band structures with and without the electric field as well as selected wavefunctions of $\Gamma$-valley and $\pm$K-valley bands are reported in Suppl. Sec. S2.

\bibliography{main.bib}

\providecommand{\noopsort}[1]{}\providecommand{\singleletter}[1]{#1}%
\begin{thebibliography}{67}%
\makeatletter
\providecommand \@ifxundefined [1]{%
 \@ifx{#1\undefined}
}%
\providecommand \@ifnum [1]{%
 \ifnum #1\expandafter \@firstoftwo
 \else \expandafter \@secondoftwo
 \fi
}%
\providecommand \@ifx [1]{%
 \ifx #1\expandafter \@firstoftwo
 \else \expandafter \@secondoftwo
 \fi
}%
\providecommand \natexlab [1]{#1}%
\providecommand \enquote  [1]{``#1''}%
\providecommand \bibnamefont  [1]{#1}%
\providecommand \bibfnamefont [1]{#1}%
\providecommand \citenamefont [1]{#1}%
\providecommand \href@noop [0]{\@secondoftwo}%
\providecommand \href [0]{\begingroup \@sanitize@url \@href}%
\providecommand \@href[1]{\@@startlink{#1}\@@href}%
\providecommand \@@href[1]{\endgroup#1\@@endlink}%
\providecommand \@sanitize@url [0]{\catcode `\\12\catcode `\$12\catcode
  `\&12\catcode `\#12\catcode `\^12\catcode `\_12\catcode `\%12\relax}%
\providecommand \@@startlink[1]{}%
\providecommand \@@endlink[0]{}%
\providecommand \url  [0]{\begingroup\@sanitize@url \@url }%
\providecommand \@url [1]{\endgroup\@href {#1}{\urlprefix }}%
\providecommand \urlprefix  [0]{URL }%
\providecommand \Eprint [0]{\href }%
\providecommand \doibase [0]{http://dx.doi.org/}%
\providecommand \selectlanguage [0]{\@gobble}%
\providecommand \bibinfo  [0]{\@secondoftwo}%
\providecommand \bibfield  [0]{\@secondoftwo}%
\providecommand \translation [1]{[#1]}%
\providecommand \BibitemOpen [0]{}%
\providecommand \bibitemStop [0]{}%
\providecommand \bibitemNoStop [0]{.\EOS\space}%
\providecommand \EOS [0]{\spacefactor3000\relax}%
\providecommand \BibitemShut  [1]{\csname bibitem#1\endcsname}%
\let\auto@bib@innerbib\@empty
\bibitem [{\citenamefont {Emery}(1987)}]{emery1987theory}%
  \BibitemOpen
  \bibfield  {author} {\bibinfo {author} {\bibfnamefont {VJ}~\bibnamefont
  {Emery}},\ }\bibfield  {title} {\enquote {\bibinfo {title} {{Theory of
  high-T$_c$ superconductivity in oxides}},}\ }\href@noop {} {\bibfield
  {journal} {\bibinfo  {journal} {Physical Review Letters}\ }\textbf {\bibinfo
  {volume} {58}},\ \bibinfo {pages} {2794} (\bibinfo {year}
  {1987})}\BibitemShut {NoStop}%
\bibitem [{\citenamefont {Zhang}\ and\ \citenamefont
  {Rice}(1988)}]{zhang1988effective}%
  \BibitemOpen
  \bibfield  {author} {\bibinfo {author} {\bibfnamefont {FC}~\bibnamefont
  {Zhang}}\ and\ \bibinfo {author} {\bibfnamefont {TM}~\bibnamefont {Rice}},\
  }\bibfield  {title} {\enquote {\bibinfo {title} {{Effective Hamiltonian for
  the superconducting Cu oxides}},}\ }\href@noop {} {\bibfield  {journal}
  {\bibinfo  {journal} {Physical Review B}\ }\textbf {\bibinfo {volume} {37}},\
  \bibinfo {pages} {3759} (\bibinfo {year} {1988})}\BibitemShut {NoStop}%
\bibitem [{\citenamefont {Dagotto}(1994)}]{dagotto1994correlated}%
  \BibitemOpen
  \bibfield  {author} {\bibinfo {author} {\bibfnamefont {Elbio}\ \bibnamefont
  {Dagotto}},\ }\bibfield  {title} {\enquote {\bibinfo {title} {Correlated
  electrons in high-temperature superconductors},}\ }\href@noop {} {\bibfield
  {journal} {\bibinfo  {journal} {Reviews of Modern Physics}\ }\textbf
  {\bibinfo {volume} {66}},\ \bibinfo {pages} {763} (\bibinfo {year}
  {1994})}\BibitemShut {NoStop}%
\bibitem [{\citenamefont {Dai}\ \emph {et~al.}(2012)\citenamefont {Dai},
  \citenamefont {Hu},\ and\ \citenamefont {Dagotto}}]{dai2012magnetism}%
  \BibitemOpen
  \bibfield  {author} {\bibinfo {author} {\bibfnamefont {Pengcheng}\
  \bibnamefont {Dai}}, \bibinfo {author} {\bibfnamefont {Jiangping}\
  \bibnamefont {Hu}}, \ and\ \bibinfo {author} {\bibfnamefont {Elbio}\
  \bibnamefont {Dagotto}},\ }\bibfield  {title} {\enquote {\bibinfo {title}
  {Magnetism and its microscopic origin in iron-based high-temperature
  superconductors},}\ }\href@noop {} {\bibfield  {journal} {\bibinfo  {journal}
  {Nature Physics}\ }\textbf {\bibinfo {volume} {8}},\ \bibinfo {pages}
  {709--718} (\bibinfo {year} {2012})}\BibitemShut {NoStop}%
\bibitem [{\citenamefont {Rohringer}\ \emph {et~al.}(2018)\citenamefont
  {Rohringer}, \citenamefont {Hafermann}, \citenamefont {Toschi}, \citenamefont
  {Katanin}, \citenamefont {Antipov}, \citenamefont {Katsnelson}, \citenamefont
  {Lichtenstein}, \citenamefont {Rubtsov},\ and\ \citenamefont
  {Held}}]{rohringer2018diagrammatic}%
  \BibitemOpen
  \bibfield  {author} {\bibinfo {author} {\bibfnamefont {G}~\bibnamefont
  {Rohringer}}, \bibinfo {author} {\bibfnamefont {H}~\bibnamefont {Hafermann}},
  \bibinfo {author} {\bibfnamefont {A}~\bibnamefont {Toschi}}, \bibinfo
  {author} {\bibfnamefont {AA}~\bibnamefont {Katanin}}, \bibinfo {author}
  {\bibfnamefont {AE}~\bibnamefont {Antipov}}, \bibinfo {author} {\bibfnamefont
  {MI}~\bibnamefont {Katsnelson}}, \bibinfo {author} {\bibfnamefont
  {AI}~\bibnamefont {Lichtenstein}}, \bibinfo {author} {\bibfnamefont
  {AN}~\bibnamefont {Rubtsov}}, \ and\ \bibinfo {author} {\bibfnamefont
  {K}~\bibnamefont {Held}},\ }\bibfield  {title} {\enquote {\bibinfo {title}
  {Diagrammatic routes to nonlocal correlations beyond dynamical mean field
  theory},}\ }\href@noop {} {\bibfield  {journal} {\bibinfo  {journal} {Reviews
  of Modern Physics}\ }\textbf {\bibinfo {volume} {90}},\ \bibinfo {pages}
  {025003} (\bibinfo {year} {2018})}\BibitemShut {NoStop}%
\bibitem [{\citenamefont {Gross}\ and\ \citenamefont
  {Bloch}(2017)}]{gross2017quantum}%
  \BibitemOpen
  \bibfield  {author} {\bibinfo {author} {\bibfnamefont {Christian}\
  \bibnamefont {Gross}}\ and\ \bibinfo {author} {\bibfnamefont {Immanuel}\
  \bibnamefont {Bloch}},\ }\bibfield  {title} {\enquote {\bibinfo {title}
  {Quantum simulations with ultracold atoms in optical lattices},}\ }\href@noop
  {} {\bibfield  {journal} {\bibinfo  {journal} {Science}\ }\textbf {\bibinfo
  {volume} {357}},\ \bibinfo {pages} {995--1001} (\bibinfo {year}
  {2017})}\BibitemShut {NoStop}%
\bibitem [{\citenamefont {Kennes}\ \emph {et~al.}(2021)\citenamefont {Kennes},
  \citenamefont {Claassen}, \citenamefont {Xian}, \citenamefont {Georges},
  \citenamefont {Millis}, \citenamefont {Hone}, \citenamefont {Dean},
  \citenamefont {Basov}, \citenamefont {Pasupathy},\ and\ \citenamefont
  {Rubio}}]{kennes2021moire}%
  \BibitemOpen
  \bibfield  {author} {\bibinfo {author} {\bibfnamefont {Dante~M}\ \bibnamefont
  {Kennes}}, \bibinfo {author} {\bibfnamefont {Martin}\ \bibnamefont
  {Claassen}}, \bibinfo {author} {\bibfnamefont {Lede}\ \bibnamefont {Xian}},
  \bibinfo {author} {\bibfnamefont {Antoine}\ \bibnamefont {Georges}}, \bibinfo
  {author} {\bibfnamefont {Andrew~J}\ \bibnamefont {Millis}}, \bibinfo {author}
  {\bibfnamefont {James}\ \bibnamefont {Hone}}, \bibinfo {author}
  {\bibfnamefont {Cory~R}\ \bibnamefont {Dean}}, \bibinfo {author}
  {\bibfnamefont {DN}~\bibnamefont {Basov}}, \bibinfo {author} {\bibfnamefont
  {Abhay~N}\ \bibnamefont {Pasupathy}}, \ and\ \bibinfo {author} {\bibfnamefont
  {Angel}\ \bibnamefont {Rubio}},\ }\bibfield  {title} {\enquote {\bibinfo
  {title} {Moir{\'e} heterostructures as a condensed-matter quantum
  simulator},}\ }\href@noop {} {\bibfield  {journal} {\bibinfo  {journal}
  {Nature Physics}\ }\textbf {\bibinfo {volume} {17}},\ \bibinfo {pages}
  {155--163} (\bibinfo {year} {2021})}\BibitemShut {NoStop}%
\bibitem [{\citenamefont {Balents}\ \emph {et~al.}(2020)\citenamefont
  {Balents}, \citenamefont {Dean}, \citenamefont {Efetov},\ and\ \citenamefont
  {Young}}]{balents2020superconductivity}%
  \BibitemOpen
  \bibfield  {author} {\bibinfo {author} {\bibfnamefont {Leon}\ \bibnamefont
  {Balents}}, \bibinfo {author} {\bibfnamefont {Cory~R}\ \bibnamefont {Dean}},
  \bibinfo {author} {\bibfnamefont {Dmitri~K}\ \bibnamefont {Efetov}}, \ and\
  \bibinfo {author} {\bibfnamefont {Andrea~F}\ \bibnamefont {Young}},\
  }\bibfield  {title} {\enquote {\bibinfo {title} {Superconductivity and strong
  correlations in moir{\'e} flat bands},}\ }\href@noop {} {\bibfield  {journal}
  {\bibinfo  {journal} {Nature Physics}\ }\textbf {\bibinfo {volume} {16}},\
  \bibinfo {pages} {725--733} (\bibinfo {year} {2020})}\BibitemShut {NoStop}%
\bibitem [{\citenamefont {Mak}\ and\ \citenamefont
  {Shan}(2022)}]{mak2022semiconductor}%
  \BibitemOpen
  \bibfield  {author} {\bibinfo {author} {\bibfnamefont {Kin~Fai}\ \bibnamefont
  {Mak}}\ and\ \bibinfo {author} {\bibfnamefont {Jie}\ \bibnamefont {Shan}},\
  }\bibfield  {title} {\enquote {\bibinfo {title} {Semiconductor moir{\'e}
  materials},}\ }\href@noop {} {\bibfield  {journal} {\bibinfo  {journal}
  {Nature Nanotechnology}\ }\textbf {\bibinfo {volume} {17}},\ \bibinfo {pages}
  {686--695} (\bibinfo {year} {2022})}\BibitemShut {NoStop}%
\bibitem [{\citenamefont {Ghiotto}\ \emph {et~al.}(2021)\citenamefont
  {Ghiotto}, \citenamefont {Shih}, \citenamefont {Pereira}, \citenamefont
  {Rhodes}, \citenamefont {Kim}, \citenamefont {Zang}, \citenamefont {Millis},
  \citenamefont {Watanabe}, \citenamefont {Taniguchi}, \citenamefont {Hone}
  \emph {et~al.}}]{ghiotto2021quantum}%
  \BibitemOpen
  \bibfield  {author} {\bibinfo {author} {\bibfnamefont {Augusto}\ \bibnamefont
  {Ghiotto}}, \bibinfo {author} {\bibfnamefont {En-Min}\ \bibnamefont {Shih}},
  \bibinfo {author} {\bibfnamefont {Giancarlo~SSG}\ \bibnamefont {Pereira}},
  \bibinfo {author} {\bibfnamefont {Daniel~A}\ \bibnamefont {Rhodes}}, \bibinfo
  {author} {\bibfnamefont {Bumho}\ \bibnamefont {Kim}}, \bibinfo {author}
  {\bibfnamefont {Jiawei}\ \bibnamefont {Zang}}, \bibinfo {author}
  {\bibfnamefont {Andrew~J}\ \bibnamefont {Millis}}, \bibinfo {author}
  {\bibfnamefont {Kenji}\ \bibnamefont {Watanabe}}, \bibinfo {author}
  {\bibfnamefont {Takashi}\ \bibnamefont {Taniguchi}}, \bibinfo {author}
  {\bibfnamefont {James~C}\ \bibnamefont {Hone}},  \emph {et~al.},\ }\bibfield
  {title} {\enquote {\bibinfo {title} {Quantum criticality in twisted
  transition metal dichalcogenides},}\ }\href@noop {} {\bibfield  {journal}
  {\bibinfo  {journal} {Nature}\ }\textbf {\bibinfo {volume} {597}},\ \bibinfo
  {pages} {345--349} (\bibinfo {year} {2021})}\BibitemShut {NoStop}%
\bibitem [{\citenamefont {Li}\ \emph {et~al.}(2021{\natexlab{a}})\citenamefont
  {Li}, \citenamefont {Jiang}, \citenamefont {Li}, \citenamefont {Zhang},
  \citenamefont {Kang}, \citenamefont {Zhu}, \citenamefont {Watanabe},
  \citenamefont {Taniguchi}, \citenamefont {Chowdhury}, \citenamefont {Fu}
  \emph {et~al.}}]{li2021continuous}%
  \BibitemOpen
  \bibfield  {author} {\bibinfo {author} {\bibfnamefont {Tingxin}\ \bibnamefont
  {Li}}, \bibinfo {author} {\bibfnamefont {Shengwei}\ \bibnamefont {Jiang}},
  \bibinfo {author} {\bibfnamefont {Lizhong}\ \bibnamefont {Li}}, \bibinfo
  {author} {\bibfnamefont {Yang}\ \bibnamefont {Zhang}}, \bibinfo {author}
  {\bibfnamefont {Kaifei}\ \bibnamefont {Kang}}, \bibinfo {author}
  {\bibfnamefont {Jiacheng}\ \bibnamefont {Zhu}}, \bibinfo {author}
  {\bibfnamefont {Kenji}\ \bibnamefont {Watanabe}}, \bibinfo {author}
  {\bibfnamefont {Takashi}\ \bibnamefont {Taniguchi}}, \bibinfo {author}
  {\bibfnamefont {Debanjan}\ \bibnamefont {Chowdhury}}, \bibinfo {author}
  {\bibfnamefont {Liang}\ \bibnamefont {Fu}},  \emph {et~al.},\ }\bibfield
  {title} {\enquote {\bibinfo {title} {Continuous mott transition in
  semiconductor moir{\'e} superlattices},}\ }\href@noop {} {\bibfield
  {journal} {\bibinfo  {journal} {Nature}\ }\textbf {\bibinfo {volume} {597}},\
  \bibinfo {pages} {350--354} (\bibinfo {year}
  {2021}{\natexlab{a}})}\BibitemShut {NoStop}%
\bibitem [{\citenamefont {Stepanov}\ \emph {et~al.}(2020)\citenamefont
  {Stepanov}, \citenamefont {Das}, \citenamefont {Lu}, \citenamefont
  {Fahimniya}, \citenamefont {Watanabe}, \citenamefont {Taniguchi},
  \citenamefont {Koppens}, \citenamefont {Lischner}, \citenamefont {Levitov},\
  and\ \citenamefont {Efetov}}]{stepanov2020untying}%
  \BibitemOpen
  \bibfield  {author} {\bibinfo {author} {\bibfnamefont {Petr}\ \bibnamefont
  {Stepanov}}, \bibinfo {author} {\bibfnamefont {Ipsita}\ \bibnamefont {Das}},
  \bibinfo {author} {\bibfnamefont {Xiaobo}\ \bibnamefont {Lu}}, \bibinfo
  {author} {\bibfnamefont {Ali}\ \bibnamefont {Fahimniya}}, \bibinfo {author}
  {\bibfnamefont {Kenji}\ \bibnamefont {Watanabe}}, \bibinfo {author}
  {\bibfnamefont {Takashi}\ \bibnamefont {Taniguchi}}, \bibinfo {author}
  {\bibfnamefont {Frank~HL}\ \bibnamefont {Koppens}}, \bibinfo {author}
  {\bibfnamefont {Johannes}\ \bibnamefont {Lischner}}, \bibinfo {author}
  {\bibfnamefont {Leonid}\ \bibnamefont {Levitov}}, \ and\ \bibinfo {author}
  {\bibfnamefont {Dmitri~K}\ \bibnamefont {Efetov}},\ }\bibfield  {title}
  {\enquote {\bibinfo {title} {Untying the insulating and superconducting
  orders in magic-angle graphene},}\ }\href@noop {} {\bibfield  {journal}
  {\bibinfo  {journal} {Nature}\ }\textbf {\bibinfo {volume} {583}},\ \bibinfo
  {pages} {375--378} (\bibinfo {year} {2020})}\BibitemShut {NoStop}%
\bibitem [{\citenamefont {Li}\ \emph {et~al.}(2021{\natexlab{b}})\citenamefont
  {Li}, \citenamefont {Zhu}, \citenamefont {Tang}, \citenamefont {Watanabe},
  \citenamefont {Taniguchi}, \citenamefont {Elser}, \citenamefont {Shan},\ and\
  \citenamefont {Mak}}]{li2021charge}%
  \BibitemOpen
  \bibfield  {author} {\bibinfo {author} {\bibfnamefont {Tingxin}\ \bibnamefont
  {Li}}, \bibinfo {author} {\bibfnamefont {Jiacheng}\ \bibnamefont {Zhu}},
  \bibinfo {author} {\bibfnamefont {Yanhao}\ \bibnamefont {Tang}}, \bibinfo
  {author} {\bibfnamefont {Kenji}\ \bibnamefont {Watanabe}}, \bibinfo {author}
  {\bibfnamefont {Takashi}\ \bibnamefont {Taniguchi}}, \bibinfo {author}
  {\bibfnamefont {Veit}\ \bibnamefont {Elser}}, \bibinfo {author}
  {\bibfnamefont {Jie}\ \bibnamefont {Shan}}, \ and\ \bibinfo {author}
  {\bibfnamefont {Kin~Fai}\ \bibnamefont {Mak}},\ }\bibfield  {title} {\enquote
  {\bibinfo {title} {Charge-order-enhanced capacitance in semiconductor
  moir{\'e} superlattices},}\ }\href@noop {} {\bibfield  {journal} {\bibinfo
  {journal} {Nature Nanotechnology}\ }\textbf {\bibinfo {volume} {16}},\
  \bibinfo {pages} {1068--1072} (\bibinfo {year}
  {2021}{\natexlab{b}})}\BibitemShut {NoStop}%
\bibitem [{\citenamefont {Wu}\ \emph {et~al.}(2018)\citenamefont {Wu},
  \citenamefont {Lovorn}, \citenamefont {Tutuc},\ and\ \citenamefont
  {MacDonald}}]{wu2018hubbard}%
  \BibitemOpen
  \bibfield  {author} {\bibinfo {author} {\bibfnamefont {Fengcheng}\
  \bibnamefont {Wu}}, \bibinfo {author} {\bibfnamefont {Timothy}\ \bibnamefont
  {Lovorn}}, \bibinfo {author} {\bibfnamefont {Emanuel}\ \bibnamefont {Tutuc}},
  \ and\ \bibinfo {author} {\bibfnamefont {Allan~H}\ \bibnamefont
  {MacDonald}},\ }\bibfield  {title} {\enquote {\bibinfo {title} {Hubbard model
  physics in transition metal dichalcogenide moir{\'e} bands},}\ }\href@noop {}
  {\bibfield  {journal} {\bibinfo  {journal} {Physical Review Letters}\
  }\textbf {\bibinfo {volume} {121}},\ \bibinfo {pages} {026402} (\bibinfo
  {year} {2018})}\BibitemShut {NoStop}%
\bibitem [{\citenamefont {Zhang}\ \emph {et~al.}(2020)\citenamefont {Zhang},
  \citenamefont {Yuan},\ and\ \citenamefont {Fu}}]{zhang2020moire}%
  \BibitemOpen
  \bibfield  {author} {\bibinfo {author} {\bibfnamefont {Yang}\ \bibnamefont
  {Zhang}}, \bibinfo {author} {\bibfnamefont {Noah~FQ}\ \bibnamefont {Yuan}}, \
  and\ \bibinfo {author} {\bibfnamefont {Liang}\ \bibnamefont {Fu}},\
  }\bibfield  {title} {\enquote {\bibinfo {title} {Moir{\'e} quantum chemistry:
  charge transfer in transition metal dichalcogenide superlattices},}\
  }\href@noop {} {\bibfield  {journal} {\bibinfo  {journal} {Physical Review
  B}\ }\textbf {\bibinfo {volume} {102}},\ \bibinfo {pages} {201115} (\bibinfo
  {year} {2020})}\BibitemShut {NoStop}%
\bibitem [{\citenamefont {Tang}\ \emph {et~al.}(2020)\citenamefont {Tang},
  \citenamefont {Li}, \citenamefont {Li}, \citenamefont {Xu}, \citenamefont
  {Liu}, \citenamefont {Barmak}, \citenamefont {Watanabe}, \citenamefont
  {Taniguchi}, \citenamefont {MacDonald}, \citenamefont {Shan} \emph
  {et~al.}}]{tang2020simulation}%
  \BibitemOpen
  \bibfield  {author} {\bibinfo {author} {\bibfnamefont {Yanhao}\ \bibnamefont
  {Tang}}, \bibinfo {author} {\bibfnamefont {Lizhong}\ \bibnamefont {Li}},
  \bibinfo {author} {\bibfnamefont {Tingxin}\ \bibnamefont {Li}}, \bibinfo
  {author} {\bibfnamefont {Yang}\ \bibnamefont {Xu}}, \bibinfo {author}
  {\bibfnamefont {Song}\ \bibnamefont {Liu}}, \bibinfo {author} {\bibfnamefont
  {Katayun}\ \bibnamefont {Barmak}}, \bibinfo {author} {\bibfnamefont {Kenji}\
  \bibnamefont {Watanabe}}, \bibinfo {author} {\bibfnamefont {Takashi}\
  \bibnamefont {Taniguchi}}, \bibinfo {author} {\bibfnamefont {Allan~H}\
  \bibnamefont {MacDonald}}, \bibinfo {author} {\bibfnamefont {Jie}\
  \bibnamefont {Shan}},  \emph {et~al.},\ }\bibfield  {title} {\enquote
  {\bibinfo {title} {{Simulation of Hubbard model physics in WSe$_2$/WS$_2$
  moir{\'e} superlattices}},}\ }\href@noop {} {\bibfield  {journal} {\bibinfo
  {journal} {Nature}\ }\textbf {\bibinfo {volume} {579}},\ \bibinfo {pages}
  {353--358} (\bibinfo {year} {2020})}\BibitemShut {NoStop}%
\bibitem [{\citenamefont {Regan}\ \emph {et~al.}(2020)\citenamefont {Regan},
  \citenamefont {Wang}, \citenamefont {Jin}, \citenamefont {Utama},
  \citenamefont {Gao}, \citenamefont {Wei}, \citenamefont {Zhao}, \citenamefont
  {Zhao}, \citenamefont {Zhang}, \citenamefont {Yumigeta} \emph
  {et~al.}}]{regan2020mott}%
  \BibitemOpen
  \bibfield  {author} {\bibinfo {author} {\bibfnamefont {Emma~C}\ \bibnamefont
  {Regan}}, \bibinfo {author} {\bibfnamefont {Danqing}\ \bibnamefont {Wang}},
  \bibinfo {author} {\bibfnamefont {Chenhao}\ \bibnamefont {Jin}}, \bibinfo
  {author} {\bibfnamefont {M~Iqbal~Bakti}\ \bibnamefont {Utama}}, \bibinfo
  {author} {\bibfnamefont {Beini}\ \bibnamefont {Gao}}, \bibinfo {author}
  {\bibfnamefont {Xin}\ \bibnamefont {Wei}}, \bibinfo {author} {\bibfnamefont
  {Sihan}\ \bibnamefont {Zhao}}, \bibinfo {author} {\bibfnamefont {Wenyu}\
  \bibnamefont {Zhao}}, \bibinfo {author} {\bibfnamefont {Zuocheng}\
  \bibnamefont {Zhang}}, \bibinfo {author} {\bibfnamefont {Kentaro}\
  \bibnamefont {Yumigeta}},  \emph {et~al.},\ }\bibfield  {title} {\enquote
  {\bibinfo {title} {{Mott and generalized Wigner crystal states in
  WSe$_2$/WS$_2$ moir{\'e} superlattices}},}\ }\href@noop {} {\bibfield
  {journal} {\bibinfo  {journal} {Nature}\ }\textbf {\bibinfo {volume} {579}},\
  \bibinfo {pages} {359--363} (\bibinfo {year} {2020})}\BibitemShut {NoStop}%
\bibitem [{\citenamefont {Shimazaki}\ \emph {et~al.}(2020)\citenamefont
  {Shimazaki}, \citenamefont {Schwartz}, \citenamefont {Watanabe},
  \citenamefont {Taniguchi}, \citenamefont {Kroner},\ and\ \citenamefont
  {Imamo{\u{g}}lu}}]{shimazaki2020strongly}%
  \BibitemOpen
  \bibfield  {author} {\bibinfo {author} {\bibfnamefont {Yuya}\ \bibnamefont
  {Shimazaki}}, \bibinfo {author} {\bibfnamefont {Ido}\ \bibnamefont
  {Schwartz}}, \bibinfo {author} {\bibfnamefont {Kenji}\ \bibnamefont
  {Watanabe}}, \bibinfo {author} {\bibfnamefont {Takashi}\ \bibnamefont
  {Taniguchi}}, \bibinfo {author} {\bibfnamefont {Martin}\ \bibnamefont
  {Kroner}}, \ and\ \bibinfo {author} {\bibfnamefont {Ata{\c{c}}}\ \bibnamefont
  {Imamo{\u{g}}lu}},\ }\bibfield  {title} {\enquote {\bibinfo {title} {Strongly
  correlated electrons and hybrid excitons in a moir{\'e} heterostructure},}\
  }\href@noop {} {\bibfield  {journal} {\bibinfo  {journal} {Nature}\ }\textbf
  {\bibinfo {volume} {580}},\ \bibinfo {pages} {472--477} (\bibinfo {year}
  {2020})}\BibitemShut {NoStop}%
\bibitem [{\citenamefont {Wang}\ \emph
  {et~al.}(2020{\natexlab{a}})\citenamefont {Wang}, \citenamefont {Shih},
  \citenamefont {Ghiotto}, \citenamefont {Xian}, \citenamefont {Rhodes},
  \citenamefont {Tan}, \citenamefont {Claassen}, \citenamefont {Kennes},
  \citenamefont {Bai}, \citenamefont {Kim} \emph
  {et~al.}}]{wang2020correlated}%
  \BibitemOpen
  \bibfield  {author} {\bibinfo {author} {\bibfnamefont {Lei}\ \bibnamefont
  {Wang}}, \bibinfo {author} {\bibfnamefont {En-Min}\ \bibnamefont {Shih}},
  \bibinfo {author} {\bibfnamefont {Augusto}\ \bibnamefont {Ghiotto}}, \bibinfo
  {author} {\bibfnamefont {Lede}\ \bibnamefont {Xian}}, \bibinfo {author}
  {\bibfnamefont {Daniel~A}\ \bibnamefont {Rhodes}}, \bibinfo {author}
  {\bibfnamefont {Cheng}\ \bibnamefont {Tan}}, \bibinfo {author} {\bibfnamefont
  {Martin}\ \bibnamefont {Claassen}}, \bibinfo {author} {\bibfnamefont
  {Dante~M}\ \bibnamefont {Kennes}}, \bibinfo {author} {\bibfnamefont {Yusong}\
  \bibnamefont {Bai}}, \bibinfo {author} {\bibfnamefont {Bumho}\ \bibnamefont
  {Kim}},  \emph {et~al.},\ }\bibfield  {title} {\enquote {\bibinfo {title}
  {Correlated electronic phases in twisted bilayer transition metal
  dichalcogenides},}\ }\href@noop {} {\bibfield  {journal} {\bibinfo  {journal}
  {Nature materials}\ }\textbf {\bibinfo {volume} {19}},\ \bibinfo {pages}
  {861--866} (\bibinfo {year} {2020}{\natexlab{a}})}\BibitemShut {NoStop}%
\bibitem [{\citenamefont {Campbell}\ \emph {et~al.}(2022)\citenamefont
  {Campbell}, \citenamefont {Brotons-Gisbert}, \citenamefont {Baek},
  \citenamefont {Vitale}, \citenamefont {Taniguchi}, \citenamefont {Watanabe},
  \citenamefont {Lischner},\ and\ \citenamefont
  {Gerardot}}]{campbell2022strongly}%
  \BibitemOpen
  \bibfield  {author} {\bibinfo {author} {\bibfnamefont {Aidan~J}\ \bibnamefont
  {Campbell}}, \bibinfo {author} {\bibfnamefont {Mauro}\ \bibnamefont
  {Brotons-Gisbert}}, \bibinfo {author} {\bibfnamefont {Hyeonjun}\ \bibnamefont
  {Baek}}, \bibinfo {author} {\bibfnamefont {Valerio}\ \bibnamefont {Vitale}},
  \bibinfo {author} {\bibfnamefont {Takashi}\ \bibnamefont {Taniguchi}},
  \bibinfo {author} {\bibfnamefont {Kenji}\ \bibnamefont {Watanabe}}, \bibinfo
  {author} {\bibfnamefont {Johannes}\ \bibnamefont {Lischner}}, \ and\ \bibinfo
  {author} {\bibfnamefont {Brian~D}\ \bibnamefont {Gerardot}},\ }\bibfield
  {title} {\enquote {\bibinfo {title} {{Exciton-polarons in the presence of
  strongly correlated electronic states in a MoSe$_2$/WSe$_2$ moir{\'e}
  superlattice}},}\ }\href@noop {} {\bibfield  {journal} {\bibinfo  {journal}
  {npj 2D Materials and Applications}\ }\textbf {\bibinfo {volume} {6}},\
  \bibinfo {pages} {1--8} (\bibinfo {year} {2022})}\BibitemShut {NoStop}%
\bibitem [{\citenamefont {Xu}\ \emph {et~al.}(2022)\citenamefont {Xu},
  \citenamefont {Kang}, \citenamefont {Watanabe}, \citenamefont {Taniguchi},
  \citenamefont {Mak},\ and\ \citenamefont {Shan}}]{xu2022tunable}%
  \BibitemOpen
  \bibfield  {author} {\bibinfo {author} {\bibfnamefont {Yang}\ \bibnamefont
  {Xu}}, \bibinfo {author} {\bibfnamefont {Kaifei}\ \bibnamefont {Kang}},
  \bibinfo {author} {\bibfnamefont {Kenji}\ \bibnamefont {Watanabe}}, \bibinfo
  {author} {\bibfnamefont {Takashi}\ \bibnamefont {Taniguchi}}, \bibinfo
  {author} {\bibfnamefont {Kin~Fai}\ \bibnamefont {Mak}}, \ and\ \bibinfo
  {author} {\bibfnamefont {Jie}\ \bibnamefont {Shan}},\ }\bibfield  {title}
  {\enquote {\bibinfo {title} {{A tunable bilayer Hubbard model in twisted
  WSe$_2$}},}\ }\href@noop {} {\bibfield  {journal} {\bibinfo  {journal}
  {Nature Nanotechnology}\ }\textbf {\bibinfo {volume} {17}},\ \bibinfo {pages}
  {934--939} (\bibinfo {year} {2022})}\BibitemShut {NoStop}%
\bibitem [{\citenamefont {Angeli}\ and\ \citenamefont
  {MacDonald}(2021)}]{angeli2021gamma}%
  \BibitemOpen
  \bibfield  {author} {\bibinfo {author} {\bibfnamefont {Mattia}\ \bibnamefont
  {Angeli}}\ and\ \bibinfo {author} {\bibfnamefont {Allan~H}\ \bibnamefont
  {MacDonald}},\ }\bibfield  {title} {\enquote {\bibinfo {title} {{$\Gamma$
  valley transition metal dichalcogenide moir{\'e} bands}},}\ }\href@noop {}
  {\bibfield  {journal} {\bibinfo  {journal} {Proceedings of the National
  Academy of Sciences}\ }\textbf {\bibinfo {volume} {118}},\ \bibinfo {pages}
  {e2021826118} (\bibinfo {year} {2021})}\BibitemShut {NoStop}%
\bibitem [{\citenamefont {Xian}\ \emph {et~al.}(2021)\citenamefont {Xian},
  \citenamefont {Claassen}, \citenamefont {Kiese}, \citenamefont {Scherer},
  \citenamefont {Trebst}, \citenamefont {Kennes},\ and\ \citenamefont
  {Rubio}}]{xian2021realization}%
  \BibitemOpen
  \bibfield  {author} {\bibinfo {author} {\bibfnamefont {Lede}\ \bibnamefont
  {Xian}}, \bibinfo {author} {\bibfnamefont {Martin}\ \bibnamefont {Claassen}},
  \bibinfo {author} {\bibfnamefont {Dominik}\ \bibnamefont {Kiese}}, \bibinfo
  {author} {\bibfnamefont {Michael~M}\ \bibnamefont {Scherer}}, \bibinfo
  {author} {\bibfnamefont {Simon}\ \bibnamefont {Trebst}}, \bibinfo {author}
  {\bibfnamefont {Dante~M}\ \bibnamefont {Kennes}}, \ and\ \bibinfo {author}
  {\bibfnamefont {Angel}\ \bibnamefont {Rubio}},\ }\bibfield  {title} {\enquote
  {\bibinfo {title} {{Realization of nearly dispersionless bands with strong
  orbital anisotropy from destructive interference in twisted bilayer
  MoS$_2$}},}\ }\href@noop {} {\bibfield  {journal} {\bibinfo  {journal}
  {Nature Communications}\ }\textbf {\bibinfo {volume} {12}} (\bibinfo {year}
  {2021})}\BibitemShut {NoStop}%
\bibitem [{\citenamefont {Kaushal}\ \emph {et~al.}(2022)\citenamefont
  {Kaushal}, \citenamefont {Morales-Dur{\'a}n}, \citenamefont {MacDonald},\
  and\ \citenamefont {Dagotto}}]{kaushal2022magnetic}%
  \BibitemOpen
  \bibfield  {author} {\bibinfo {author} {\bibfnamefont {Nitin}\ \bibnamefont
  {Kaushal}}, \bibinfo {author} {\bibfnamefont {Nicol{\'a}s}\ \bibnamefont
  {Morales-Dur{\'a}n}}, \bibinfo {author} {\bibfnamefont {Allan~H}\
  \bibnamefont {MacDonald}}, \ and\ \bibinfo {author} {\bibfnamefont {Elbio}\
  \bibnamefont {Dagotto}},\ }\bibfield  {title} {\enquote {\bibinfo {title}
  {{Magnetic ground states of honeycomb lattice Wigner crystals}},}\
  }\href@noop {} {\bibfield  {journal} {\bibinfo  {journal} {Communications
  Physics}\ }\textbf {\bibinfo {volume} {5}},\ \bibinfo {pages} {289} (\bibinfo
  {year} {2022})}\BibitemShut {NoStop}%
\bibitem [{\citenamefont {An}\ \emph {et~al.}(2020)\citenamefont {An},
  \citenamefont {Cai}, \citenamefont {Pei}, \citenamefont {Huang},
  \citenamefont {Wu}, \citenamefont {Zhou}, \citenamefont {Lin}, \citenamefont
  {Ying}, \citenamefont {Ye}, \citenamefont {Feng} \emph
  {et~al.}}]{an2020interaction}%
  \BibitemOpen
  \bibfield  {author} {\bibinfo {author} {\bibfnamefont {Liheng}\ \bibnamefont
  {An}}, \bibinfo {author} {\bibfnamefont {Xiangbin}\ \bibnamefont {Cai}},
  \bibinfo {author} {\bibfnamefont {Ding}\ \bibnamefont {Pei}}, \bibinfo
  {author} {\bibfnamefont {Meizhen}\ \bibnamefont {Huang}}, \bibinfo {author}
  {\bibfnamefont {Zefei}\ \bibnamefont {Wu}}, \bibinfo {author} {\bibfnamefont
  {Zishu}\ \bibnamefont {Zhou}}, \bibinfo {author} {\bibfnamefont {Jiangxiazi}\
  \bibnamefont {Lin}}, \bibinfo {author} {\bibfnamefont {Zhehan}\ \bibnamefont
  {Ying}}, \bibinfo {author} {\bibfnamefont {Ziqing}\ \bibnamefont {Ye}},
  \bibinfo {author} {\bibfnamefont {Xuemeng}\ \bibnamefont {Feng}},  \emph
  {et~al.},\ }\bibfield  {title} {\enquote {\bibinfo {title} {{Interaction
  effects and superconductivity signatures in twisted double-bilayer
  WSe$_2$}},}\ }\href@noop {} {\bibfield  {journal} {\bibinfo  {journal}
  {Nanoscale Horizons}\ }\textbf {\bibinfo {volume} {5}},\ \bibinfo {pages}
  {1309--1316} (\bibinfo {year} {2020})}\BibitemShut {NoStop}%
\bibitem [{\citenamefont {Pei}\ \emph {et~al.}(2022)\citenamefont {Pei},
  \citenamefont {Wang}, \citenamefont {Zhou}, \citenamefont {He}, \citenamefont
  {An}, \citenamefont {He}, \citenamefont {Chen}, \citenamefont {Li},
  \citenamefont {Wei}, \citenamefont {Liang} \emph
  {et~al.}}]{pei2022observation}%
  \BibitemOpen
  \bibfield  {author} {\bibinfo {author} {\bibfnamefont {Ding}\ \bibnamefont
  {Pei}}, \bibinfo {author} {\bibfnamefont {Binbin}\ \bibnamefont {Wang}},
  \bibinfo {author} {\bibfnamefont {Zishu}\ \bibnamefont {Zhou}}, \bibinfo
  {author} {\bibfnamefont {Zhihai}\ \bibnamefont {He}}, \bibinfo {author}
  {\bibfnamefont {Liheng}\ \bibnamefont {An}}, \bibinfo {author} {\bibfnamefont
  {Shanmei}\ \bibnamefont {He}}, \bibinfo {author} {\bibfnamefont {Cheng}\
  \bibnamefont {Chen}}, \bibinfo {author} {\bibfnamefont {Yiwei}\ \bibnamefont
  {Li}}, \bibinfo {author} {\bibfnamefont {Liyang}\ \bibnamefont {Wei}},
  \bibinfo {author} {\bibfnamefont {Aiji}\ \bibnamefont {Liang}},  \emph
  {et~al.},\ }\bibfield  {title} {\enquote {\bibinfo {title} {{Observation of
  $\Gamma$-valley moir{\'e} bands and emergent hexagonal lattice in twisted
  transition metal dichalcogenides}},}\ }\href@noop {} {\bibfield  {journal}
  {\bibinfo  {journal} {Physical Review X}\ ,\ \bibinfo {pages} {021065}}
  (\bibinfo {year} {2022})}\BibitemShut {NoStop}%
\bibitem [{\citenamefont {Gatti}\ \emph {et~al.}(2022)\citenamefont {Gatti},
  \citenamefont {Issing}, \citenamefont {Rademaker}, \citenamefont {Margot},
  \citenamefont {de~Jong}, \citenamefont {van~der Molen}, \citenamefont
  {Teyssier}, \citenamefont {Kim}, \citenamefont {Watson}, \citenamefont
  {Cacho} \emph {et~al.}}]{gatti2022observation}%
  \BibitemOpen
  \bibfield  {author} {\bibinfo {author} {\bibfnamefont {Gianmarco}\
  \bibnamefont {Gatti}}, \bibinfo {author} {\bibfnamefont {Julia}\ \bibnamefont
  {Issing}}, \bibinfo {author} {\bibfnamefont {Louk}\ \bibnamefont
  {Rademaker}}, \bibinfo {author} {\bibfnamefont {Florian}\ \bibnamefont
  {Margot}}, \bibinfo {author} {\bibfnamefont {Tobias~A}\ \bibnamefont
  {de~Jong}}, \bibinfo {author} {\bibfnamefont {Sense~Jan}\ \bibnamefont
  {van~der Molen}}, \bibinfo {author} {\bibfnamefont {J{\'e}r{\'e}mie}\
  \bibnamefont {Teyssier}}, \bibinfo {author} {\bibfnamefont {Timur~K}\
  \bibnamefont {Kim}}, \bibinfo {author} {\bibfnamefont {Matthew~D}\
  \bibnamefont {Watson}}, \bibinfo {author} {\bibfnamefont {Cephise}\
  \bibnamefont {Cacho}},  \emph {et~al.},\ }\bibfield  {title} {\enquote
  {\bibinfo {title} {{Observation of flat $\Gamma $ moir{\'e} bands in twisted
  bilayer WSe$_2$}},}\ }\href@noop {} {\bibfield  {journal} {\bibinfo
  {journal} {arXiv preprint arXiv:2211.01192}\ } (\bibinfo {year}
  {2022})}\BibitemShut {NoStop}%
\bibitem [{\citenamefont {Foutty}\ \emph {et~al.}(2022)\citenamefont {Foutty},
  \citenamefont {Yu}, \citenamefont {Devakul}, \citenamefont {Kometter},
  \citenamefont {Zhang}, \citenamefont {Watanabe}, \citenamefont {Taniguchi},
  \citenamefont {Fu},\ and\ \citenamefont {Feldman}}]{foutty2022tunable}%
  \BibitemOpen
  \bibfield  {author} {\bibinfo {author} {\bibfnamefont {Benjamin~A}\
  \bibnamefont {Foutty}}, \bibinfo {author} {\bibfnamefont {Jiachen}\
  \bibnamefont {Yu}}, \bibinfo {author} {\bibfnamefont {Trithep}\ \bibnamefont
  {Devakul}}, \bibinfo {author} {\bibfnamefont {Carlos~R}\ \bibnamefont
  {Kometter}}, \bibinfo {author} {\bibfnamefont {Yang}\ \bibnamefont {Zhang}},
  \bibinfo {author} {\bibfnamefont {Kenji}\ \bibnamefont {Watanabe}}, \bibinfo
  {author} {\bibfnamefont {Takashi}\ \bibnamefont {Taniguchi}}, \bibinfo
  {author} {\bibfnamefont {Liang}\ \bibnamefont {Fu}}, \ and\ \bibinfo {author}
  {\bibfnamefont {Benjamin~E}\ \bibnamefont {Feldman}},\ }\bibfield  {title}
  {\enquote {\bibinfo {title} {{Tunable spin and valley excitations of
  correlated insulators in $\Gamma$-valley moir{\'e} bands}},}\ }\href@noop {}
  {\bibfield  {journal} {\bibinfo  {journal} {arXiv preprint arXiv:2206.10631}\
  } (\bibinfo {year} {2022})}\BibitemShut {NoStop}%
\bibitem [{\citenamefont {Vitale}\ \emph {et~al.}(2021)\citenamefont {Vitale},
  \citenamefont {Atalar}, \citenamefont {Mostofi},\ and\ \citenamefont
  {Lischner}}]{vitale2021flat}%
  \BibitemOpen
  \bibfield  {author} {\bibinfo {author} {\bibfnamefont {Valerio}\ \bibnamefont
  {Vitale}}, \bibinfo {author} {\bibfnamefont {Kemal}\ \bibnamefont {Atalar}},
  \bibinfo {author} {\bibfnamefont {Arash~A}\ \bibnamefont {Mostofi}}, \ and\
  \bibinfo {author} {\bibfnamefont {Johannes}\ \bibnamefont {Lischner}},\
  }\bibfield  {title} {\enquote {\bibinfo {title} {{Flat band properties of
  twisted transition metal dichalcogenide homo-and heterobilayers of MoS$_2$,
  MoSe$_2$, WS$_2$ and WSe$_2$}},}\ }\href@noop {} {\bibfield  {journal}
  {\bibinfo  {journal} {{2D Materials}}\ }\textbf {\bibinfo {volume} {8}},\
  \bibinfo {pages} {045010} (\bibinfo {year} {2021})}\BibitemShut {NoStop}%
\bibitem [{\citenamefont {Back}\ \emph {et~al.}(2017)\citenamefont {Back},
  \citenamefont {Sidler}, \citenamefont {Cotlet}, \citenamefont {Srivastava},
  \citenamefont {Takemura}, \citenamefont {Kroner},\ and\ \citenamefont
  {Imamo{\u{g}}lu}}]{back2017giant}%
  \BibitemOpen
  \bibfield  {author} {\bibinfo {author} {\bibfnamefont {Patrick}\ \bibnamefont
  {Back}}, \bibinfo {author} {\bibfnamefont {Meinrad}\ \bibnamefont {Sidler}},
  \bibinfo {author} {\bibfnamefont {Ovidiu}\ \bibnamefont {Cotlet}}, \bibinfo
  {author} {\bibfnamefont {Ajit}\ \bibnamefont {Srivastava}}, \bibinfo {author}
  {\bibfnamefont {Naotomo}\ \bibnamefont {Takemura}}, \bibinfo {author}
  {\bibfnamefont {Martin}\ \bibnamefont {Kroner}}, \ and\ \bibinfo {author}
  {\bibfnamefont {Atac}\ \bibnamefont {Imamo{\u{g}}lu}},\ }\bibfield  {title}
  {\enquote {\bibinfo {title} {{Giant paramagnetism-induced valley polarization
  of electrons in charge-tunable monolayer MoSe$_2$}},}\ }\href@noop {}
  {\bibfield  {journal} {\bibinfo  {journal} {Physical Review Letters}\
  }\textbf {\bibinfo {volume} {118}},\ \bibinfo {pages} {237404} (\bibinfo
  {year} {2017})}\BibitemShut {NoStop}%
\bibitem [{\citenamefont {Roch}\ \emph {et~al.}(2019)\citenamefont {Roch},
  \citenamefont {Froehlicher}, \citenamefont {Leisgang}, \citenamefont {Makk},
  \citenamefont {Watanabe}, \citenamefont {Taniguchi},\ and\ \citenamefont
  {Warburton}}]{roch2019spin}%
  \BibitemOpen
  \bibfield  {author} {\bibinfo {author} {\bibfnamefont {Jonas~Ga{\"e}l}\
  \bibnamefont {Roch}}, \bibinfo {author} {\bibfnamefont {Guillaume}\
  \bibnamefont {Froehlicher}}, \bibinfo {author} {\bibfnamefont {Nadine}\
  \bibnamefont {Leisgang}}, \bibinfo {author} {\bibfnamefont {Peter}\
  \bibnamefont {Makk}}, \bibinfo {author} {\bibfnamefont {Kenji}\ \bibnamefont
  {Watanabe}}, \bibinfo {author} {\bibfnamefont {Takashi}\ \bibnamefont
  {Taniguchi}}, \ and\ \bibinfo {author} {\bibfnamefont {Richard~John}\
  \bibnamefont {Warburton}},\ }\bibfield  {title} {\enquote {\bibinfo {title}
  {{Spin-polarized electrons in monolayer MoS$_2$}},}\ }\href@noop {}
  {\bibfield  {journal} {\bibinfo  {journal} {Nature Nanotechnology}\ }\textbf
  {\bibinfo {volume} {14}},\ \bibinfo {pages} {432--436} (\bibinfo {year}
  {2019})}\BibitemShut {NoStop}%
\bibitem [{\citenamefont {Liu}\ \emph {et~al.}(2021)\citenamefont {Liu},
  \citenamefont {van Baren}, \citenamefont {Lu}, \citenamefont {Taniguchi},
  \citenamefont {Watanabe}, \citenamefont {Smirnov}, \citenamefont {Chang},\
  and\ \citenamefont {Lui}}]{liu2021exciton}%
  \BibitemOpen
  \bibfield  {author} {\bibinfo {author} {\bibfnamefont {Erfu}\ \bibnamefont
  {Liu}}, \bibinfo {author} {\bibfnamefont {Jeremiah}\ \bibnamefont {van
  Baren}}, \bibinfo {author} {\bibfnamefont {Zhengguang}\ \bibnamefont {Lu}},
  \bibinfo {author} {\bibfnamefont {Takashi}\ \bibnamefont {Taniguchi}},
  \bibinfo {author} {\bibfnamefont {Kenji}\ \bibnamefont {Watanabe}}, \bibinfo
  {author} {\bibfnamefont {Dmitry}\ \bibnamefont {Smirnov}}, \bibinfo {author}
  {\bibfnamefont {Yia-Chung}\ \bibnamefont {Chang}}, \ and\ \bibinfo {author}
  {\bibfnamefont {Chun~Hung}\ \bibnamefont {Lui}},\ }\bibfield  {title}
  {\enquote {\bibinfo {title} {{Exciton-polaron Rydberg states in monolayer
  MoSe$_2$ and WSe$_2$}},}\ }\href@noop {} {\bibfield  {journal} {\bibinfo
  {journal} {Nature Communications}\ }\textbf {\bibinfo {volume} {12}}
  (\bibinfo {year} {2021})}\BibitemShut {NoStop}%
\bibitem [{\citenamefont {Zhou}\ \emph {et~al.}(2021)\citenamefont {Zhou},
  \citenamefont {Sung}, \citenamefont {Brutschea}, \citenamefont {Esterlis},
  \citenamefont {Wang}, \citenamefont {Scuri}, \citenamefont {Gelly},
  \citenamefont {Heo}, \citenamefont {Taniguchi}, \citenamefont {Watanabe}
  \emph {et~al.}}]{zhou2021bilayer}%
  \BibitemOpen
  \bibfield  {author} {\bibinfo {author} {\bibfnamefont {You}\ \bibnamefont
  {Zhou}}, \bibinfo {author} {\bibfnamefont {Jiho}\ \bibnamefont {Sung}},
  \bibinfo {author} {\bibfnamefont {Elise}\ \bibnamefont {Brutschea}}, \bibinfo
  {author} {\bibfnamefont {Ilya}\ \bibnamefont {Esterlis}}, \bibinfo {author}
  {\bibfnamefont {Yao}\ \bibnamefont {Wang}}, \bibinfo {author} {\bibfnamefont
  {Giovanni}\ \bibnamefont {Scuri}}, \bibinfo {author} {\bibfnamefont {Ryan~J}\
  \bibnamefont {Gelly}}, \bibinfo {author} {\bibfnamefont {Hoseok}\
  \bibnamefont {Heo}}, \bibinfo {author} {\bibfnamefont {Takashi}\ \bibnamefont
  {Taniguchi}}, \bibinfo {author} {\bibfnamefont {Kenji}\ \bibnamefont
  {Watanabe}},  \emph {et~al.},\ }\bibfield  {title} {\enquote {\bibinfo
  {title} {{Bilayer Wigner crystals in a transition metal dichalcogenide
  heterostructure}},}\ }\href@noop {} {\bibfield  {journal} {\bibinfo
  {journal} {Nature}\ }\textbf {\bibinfo {volume} {595}},\ \bibinfo {pages}
  {48--52} (\bibinfo {year} {2021})}\BibitemShut {NoStop}%
\bibitem [{\citenamefont {Movva}\ \emph {et~al.}(2018)\citenamefont {Movva},
  \citenamefont {Lovorn}, \citenamefont {Fallahazad}, \citenamefont {Larentis},
  \citenamefont {Kim}, \citenamefont {Taniguchi}, \citenamefont {Watanabe},
  \citenamefont {Banerjee}, \citenamefont {MacDonald},\ and\ \citenamefont
  {Tutuc}}]{movva2018tunable}%
  \BibitemOpen
  \bibfield  {author} {\bibinfo {author} {\bibfnamefont {Hema~CP}\ \bibnamefont
  {Movva}}, \bibinfo {author} {\bibfnamefont {Timothy}\ \bibnamefont {Lovorn}},
  \bibinfo {author} {\bibfnamefont {Babak}\ \bibnamefont {Fallahazad}},
  \bibinfo {author} {\bibfnamefont {Stefano}\ \bibnamefont {Larentis}},
  \bibinfo {author} {\bibfnamefont {Kyounghwan}\ \bibnamefont {Kim}}, \bibinfo
  {author} {\bibfnamefont {Takashi}\ \bibnamefont {Taniguchi}}, \bibinfo
  {author} {\bibfnamefont {Kenji}\ \bibnamefont {Watanabe}}, \bibinfo {author}
  {\bibfnamefont {Sanjay~K}\ \bibnamefont {Banerjee}}, \bibinfo {author}
  {\bibfnamefont {Allan~H}\ \bibnamefont {MacDonald}}, \ and\ \bibinfo {author}
  {\bibfnamefont {Emanuel}\ \bibnamefont {Tutuc}},\ }\bibfield  {title}
  {\enquote {\bibinfo {title} {{Tunable $\Gamma$- K Valley Populations in
  Hole-Doped Trilayer WSe$_2$}},}\ }\href@noop {} {\bibfield  {journal}
  {\bibinfo  {journal} {Physical Review Letters}\ }\textbf {\bibinfo {volume}
  {120}},\ \bibinfo {pages} {107703} (\bibinfo {year} {2018})}\BibitemShut
  {NoStop}%
\bibitem [{\citenamefont {Jin}\ \emph {et~al.}(2019)\citenamefont {Jin},
  \citenamefont {Regan}, \citenamefont {Yan}, \citenamefont {Utama},
  \citenamefont {Wang}, \citenamefont {Zhao}, \citenamefont {Qin},
  \citenamefont {Yang}, \citenamefont {Zheng}, \citenamefont {Shi} \emph
  {et~al.}}]{jin2019observation}%
  \BibitemOpen
  \bibfield  {author} {\bibinfo {author} {\bibfnamefont {Chenhao}\ \bibnamefont
  {Jin}}, \bibinfo {author} {\bibfnamefont {Emma~C}\ \bibnamefont {Regan}},
  \bibinfo {author} {\bibfnamefont {Aiming}\ \bibnamefont {Yan}}, \bibinfo
  {author} {\bibfnamefont {M~Iqbal~Bakti}\ \bibnamefont {Utama}}, \bibinfo
  {author} {\bibfnamefont {Danqing}\ \bibnamefont {Wang}}, \bibinfo {author}
  {\bibfnamefont {Sihan}\ \bibnamefont {Zhao}}, \bibinfo {author}
  {\bibfnamefont {Ying}\ \bibnamefont {Qin}}, \bibinfo {author} {\bibfnamefont
  {Sijie}\ \bibnamefont {Yang}}, \bibinfo {author} {\bibfnamefont {Zhiren}\
  \bibnamefont {Zheng}}, \bibinfo {author} {\bibfnamefont {Shenyang}\
  \bibnamefont {Shi}},  \emph {et~al.},\ }\bibfield  {title} {\enquote
  {\bibinfo {title} {{Observation of moir{\'e} excitons in WSe$_2$/WS$_2$
  heterostructure superlattices}},}\ }\href@noop {} {\bibfield  {journal}
  {\bibinfo  {journal} {Nature}\ }\textbf {\bibinfo {volume} {567}},\ \bibinfo
  {pages} {76--80} (\bibinfo {year} {2019})}\BibitemShut {NoStop}%
\bibitem [{\citenamefont {Ruiz-Tijerina}\ and\ \citenamefont
  {Fal'ko}(2019)}]{ruiz2019interlayer}%
  \BibitemOpen
  \bibfield  {author} {\bibinfo {author} {\bibfnamefont {David~A}\ \bibnamefont
  {Ruiz-Tijerina}}\ and\ \bibinfo {author} {\bibfnamefont {Vladimir~I}\
  \bibnamefont {Fal'ko}},\ }\bibfield  {title} {\enquote {\bibinfo {title}
  {Interlayer hybridization and moir{\'e} superlattice minibands for electrons
  and excitons in heterobilayers of transition-metal dichalcogenides},}\
  }\href@noop {} {\bibfield  {journal} {\bibinfo  {journal} {Physical Review
  B}\ }\textbf {\bibinfo {volume} {99}},\ \bibinfo {pages} {125424} (\bibinfo
  {year} {2019})}\BibitemShut {NoStop}%
\bibitem [{\citenamefont {Jones}\ \emph {et~al.}(2014)\citenamefont {Jones},
  \citenamefont {Yu}, \citenamefont {Ross}, \citenamefont {Klement},
  \citenamefont {Ghimire}, \citenamefont {Yan}, \citenamefont {Mandrus},
  \citenamefont {Yao},\ and\ \citenamefont {Xu}}]{jones2014spin}%
  \BibitemOpen
  \bibfield  {author} {\bibinfo {author} {\bibfnamefont {Aaron~M}\ \bibnamefont
  {Jones}}, \bibinfo {author} {\bibfnamefont {Hongyi}\ \bibnamefont {Yu}},
  \bibinfo {author} {\bibfnamefont {Jason~S}\ \bibnamefont {Ross}}, \bibinfo
  {author} {\bibfnamefont {Philip}\ \bibnamefont {Klement}}, \bibinfo {author}
  {\bibfnamefont {Nirmal~J}\ \bibnamefont {Ghimire}}, \bibinfo {author}
  {\bibfnamefont {Jiaqiang}\ \bibnamefont {Yan}}, \bibinfo {author}
  {\bibfnamefont {David~G}\ \bibnamefont {Mandrus}}, \bibinfo {author}
  {\bibfnamefont {Wang}\ \bibnamefont {Yao}}, \ and\ \bibinfo {author}
  {\bibfnamefont {Xiaodong}\ \bibnamefont {Xu}},\ }\bibfield  {title} {\enquote
  {\bibinfo {title} {{Spin--layer locking effects in optical orientation of
  exciton spin in bilayer WSe$_2$}},}\ }\href@noop {} {\bibfield  {journal}
  {\bibinfo  {journal} {Nature Physics}\ }\textbf {\bibinfo {volume} {10}},\
  \bibinfo {pages} {130--134} (\bibinfo {year} {2014})}\BibitemShut {NoStop}%
\bibitem [{\citenamefont {Brotons-Gisbert}\ \emph {et~al.}(2020)\citenamefont
  {Brotons-Gisbert}, \citenamefont {Baek}, \citenamefont {Molina-S{\'a}nchez},
  \citenamefont {Campbell}, \citenamefont {Scerri}, \citenamefont {White},
  \citenamefont {Watanabe}, \citenamefont {Taniguchi}, \citenamefont {Bonato},\
  and\ \citenamefont {Gerardot}}]{brotons2020spin}%
  \BibitemOpen
  \bibfield  {author} {\bibinfo {author} {\bibfnamefont {Mauro}\ \bibnamefont
  {Brotons-Gisbert}}, \bibinfo {author} {\bibfnamefont {Hyeonjun}\ \bibnamefont
  {Baek}}, \bibinfo {author} {\bibfnamefont {Alejandro}\ \bibnamefont
  {Molina-S{\'a}nchez}}, \bibinfo {author} {\bibfnamefont {Aidan}\ \bibnamefont
  {Campbell}}, \bibinfo {author} {\bibfnamefont {Eleanor}\ \bibnamefont
  {Scerri}}, \bibinfo {author} {\bibfnamefont {Daniel}\ \bibnamefont {White}},
  \bibinfo {author} {\bibfnamefont {Kenji}\ \bibnamefont {Watanabe}}, \bibinfo
  {author} {\bibfnamefont {Takashi}\ \bibnamefont {Taniguchi}}, \bibinfo
  {author} {\bibfnamefont {Cristian}\ \bibnamefont {Bonato}}, \ and\ \bibinfo
  {author} {\bibfnamefont {Brian~D}\ \bibnamefont {Gerardot}},\ }\bibfield
  {title} {\enquote {\bibinfo {title} {Spin-layer locking of interlayer
  excitons trapped in moir{\'e} potentials},}\ }\href@noop {} {\bibfield
  {journal} {\bibinfo  {journal} {Nature Materials}\ }\textbf {\bibinfo
  {volume} {19}},\ \bibinfo {pages} {630--636} (\bibinfo {year}
  {2020})}\BibitemShut {NoStop}%
\bibitem [{\citenamefont {Courtade}\ \emph {et~al.}(2017)\citenamefont
  {Courtade}, \citenamefont {Semina}, \citenamefont {Manca}, \citenamefont
  {Glazov}, \citenamefont {Robert}, \citenamefont {Cadiz}, \citenamefont
  {Wang}, \citenamefont {Taniguchi}, \citenamefont {Watanabe}, \citenamefont
  {Pierre} \emph {et~al.}}]{courtade2017charged}%
  \BibitemOpen
  \bibfield  {author} {\bibinfo {author} {\bibfnamefont {E}~\bibnamefont
  {Courtade}}, \bibinfo {author} {\bibfnamefont {M}~\bibnamefont {Semina}},
  \bibinfo {author} {\bibfnamefont {M}~\bibnamefont {Manca}}, \bibinfo {author}
  {\bibfnamefont {MM}~\bibnamefont {Glazov}}, \bibinfo {author} {\bibfnamefont
  {C{\'e}dric}\ \bibnamefont {Robert}}, \bibinfo {author} {\bibfnamefont
  {F}~\bibnamefont {Cadiz}}, \bibinfo {author} {\bibfnamefont {G}~\bibnamefont
  {Wang}}, \bibinfo {author} {\bibfnamefont {T}~\bibnamefont {Taniguchi}},
  \bibinfo {author} {\bibfnamefont {K}~\bibnamefont {Watanabe}}, \bibinfo
  {author} {\bibfnamefont {M}~\bibnamefont {Pierre}},  \emph {et~al.},\
  }\bibfield  {title} {\enquote {\bibinfo {title} {{Charged excitons in
  monolayer WSe$_2$: experiment and theory}},}\ }\href@noop {} {\bibfield
  {journal} {\bibinfo  {journal} {Physical Review B}\ }\textbf {\bibinfo
  {volume} {96}},\ \bibinfo {pages} {085302} (\bibinfo {year}
  {2017})}\BibitemShut {NoStop}%
\bibitem [{\citenamefont {Wang}\ \emph {et~al.}(2018)\citenamefont {Wang},
  \citenamefont {Chiu}, \citenamefont {Honz}, \citenamefont {Mak},\ and\
  \citenamefont {Shan}}]{wang2018electrical}%
  \BibitemOpen
  \bibfield  {author} {\bibinfo {author} {\bibfnamefont {Zefang}\ \bibnamefont
  {Wang}}, \bibinfo {author} {\bibfnamefont {Yi-Hsin}\ \bibnamefont {Chiu}},
  \bibinfo {author} {\bibfnamefont {Kevin}\ \bibnamefont {Honz}}, \bibinfo
  {author} {\bibfnamefont {Kin~Fai}\ \bibnamefont {Mak}}, \ and\ \bibinfo
  {author} {\bibfnamefont {Jie}\ \bibnamefont {Shan}},\ }\bibfield  {title}
  {\enquote {\bibinfo {title} {{Electrical tuning of interlayer exciton gases
  in WSe$_2$ bilayers}},}\ }\href@noop {} {\bibfield  {journal} {\bibinfo
  {journal} {{Nano Letters}}\ }\textbf {\bibinfo {volume} {18}},\ \bibinfo
  {pages} {137--143} (\bibinfo {year} {2018})}\BibitemShut {NoStop}%
\bibitem [{\citenamefont {Naik}\ \emph {et~al.}(2022)\citenamefont {Naik},
  \citenamefont {Regan}, \citenamefont {Zhang}, \citenamefont {Chan},
  \citenamefont {Li}, \citenamefont {Wang}, \citenamefont {Yoon}, \citenamefont
  {Ong}, \citenamefont {Zhao}, \citenamefont {Zhao} \emph
  {et~al.}}]{naik2022nature}%
  \BibitemOpen
  \bibfield  {author} {\bibinfo {author} {\bibfnamefont {Mit~H}\ \bibnamefont
  {Naik}}, \bibinfo {author} {\bibfnamefont {Emma~C}\ \bibnamefont {Regan}},
  \bibinfo {author} {\bibfnamefont {Zuocheng}\ \bibnamefont {Zhang}}, \bibinfo
  {author} {\bibfnamefont {Yang-Hao}\ \bibnamefont {Chan}}, \bibinfo {author}
  {\bibfnamefont {Zhenglu}\ \bibnamefont {Li}}, \bibinfo {author}
  {\bibfnamefont {Danqing}\ \bibnamefont {Wang}}, \bibinfo {author}
  {\bibfnamefont {Yoseob}\ \bibnamefont {Yoon}}, \bibinfo {author}
  {\bibfnamefont {Chin~Shen}\ \bibnamefont {Ong}}, \bibinfo {author}
  {\bibfnamefont {Wenyu}\ \bibnamefont {Zhao}}, \bibinfo {author}
  {\bibfnamefont {Sihan}\ \bibnamefont {Zhao}},  \emph {et~al.},\ }\bibfield
  {title} {\enquote {\bibinfo {title} {{Intralayer charge-transfer moir{\'e}
  excitons in van der Waals superlattices}},}\ }\href@noop {} {\bibfield
  {journal} {\bibinfo  {journal} {Nature}\ }\textbf {\bibinfo {volume} {609}},\
  \bibinfo {pages} {52--57} (\bibinfo {year} {2022})}\BibitemShut {NoStop}%
\bibitem [{\citenamefont {Liu}\ \emph {et~al.}(2020)\citenamefont {Liu},
  \citenamefont {van Baren}, \citenamefont {Taniguchi}, \citenamefont
  {Watanabe}, \citenamefont {Chang},\ and\ \citenamefont
  {Lui}}]{liu2020landau}%
  \BibitemOpen
  \bibfield  {author} {\bibinfo {author} {\bibfnamefont {Erfu}\ \bibnamefont
  {Liu}}, \bibinfo {author} {\bibfnamefont {Jeremiah}\ \bibnamefont {van
  Baren}}, \bibinfo {author} {\bibfnamefont {Takashi}\ \bibnamefont
  {Taniguchi}}, \bibinfo {author} {\bibfnamefont {Kenji}\ \bibnamefont
  {Watanabe}}, \bibinfo {author} {\bibfnamefont {Yia-Chung}\ \bibnamefont
  {Chang}}, \ and\ \bibinfo {author} {\bibfnamefont {Chun~Hung}\ \bibnamefont
  {Lui}},\ }\bibfield  {title} {\enquote {\bibinfo {title} {Landau-quantized
  excitonic absorption and luminescence in a monolayer valley semiconductor},}\
  }\href@noop {} {\bibfield  {journal} {\bibinfo  {journal} {Physical Review
  Letters}\ }\textbf {\bibinfo {volume} {124}},\ \bibinfo {pages} {097401}
  (\bibinfo {year} {2020})}\BibitemShut {NoStop}%
\bibitem [{\citenamefont {Wang}\ \emph
  {et~al.}(2020{\natexlab{b}})\citenamefont {Wang}, \citenamefont {Li},
  \citenamefont {Lu}, \citenamefont {Li}, \citenamefont {Miao}, \citenamefont
  {Lian}, \citenamefont {Meng}, \citenamefont {Blei}, \citenamefont
  {Taniguchi}, \citenamefont {Watanabe} \emph {et~al.}}]{wang2020observation}%
  \BibitemOpen
  \bibfield  {author} {\bibinfo {author} {\bibfnamefont {Tianmeng}\
  \bibnamefont {Wang}}, \bibinfo {author} {\bibfnamefont {Zhipeng}\
  \bibnamefont {Li}}, \bibinfo {author} {\bibfnamefont {Zhengguang}\
  \bibnamefont {Lu}}, \bibinfo {author} {\bibfnamefont {Yunmei}\ \bibnamefont
  {Li}}, \bibinfo {author} {\bibfnamefont {Shengnan}\ \bibnamefont {Miao}},
  \bibinfo {author} {\bibfnamefont {Zhen}\ \bibnamefont {Lian}}, \bibinfo
  {author} {\bibfnamefont {Yuze}\ \bibnamefont {Meng}}, \bibinfo {author}
  {\bibfnamefont {Mark}\ \bibnamefont {Blei}}, \bibinfo {author} {\bibfnamefont
  {Takashi}\ \bibnamefont {Taniguchi}}, \bibinfo {author} {\bibfnamefont
  {Kenji}\ \bibnamefont {Watanabe}},  \emph {et~al.},\ }\bibfield  {title}
  {\enquote {\bibinfo {title} {{Observation of Quantized Exciton Energies in
  Monolayer WSe$_2$ under a Strong Magnetic Field}},}\ }\href@noop {}
  {\bibfield  {journal} {\bibinfo  {journal} {Physical Review X}\ }\textbf
  {\bibinfo {volume} {10}},\ \bibinfo {pages} {021024} (\bibinfo {year}
  {2020}{\natexlab{b}})}\BibitemShut {NoStop}%
\bibitem [{\citenamefont {Srivastava}\ \emph {et~al.}(2015)\citenamefont
  {Srivastava}, \citenamefont {Sidler}, \citenamefont {Allain}, \citenamefont
  {Lembke}, \citenamefont {Kis},\ and\ \citenamefont
  {Imamo{\u{g}}lu}}]{srivastava2015valley}%
  \BibitemOpen
  \bibfield  {author} {\bibinfo {author} {\bibfnamefont {Ajit}\ \bibnamefont
  {Srivastava}}, \bibinfo {author} {\bibfnamefont {Meinrad}\ \bibnamefont
  {Sidler}}, \bibinfo {author} {\bibfnamefont {Adrien~V}\ \bibnamefont
  {Allain}}, \bibinfo {author} {\bibfnamefont {Dominik~S}\ \bibnamefont
  {Lembke}}, \bibinfo {author} {\bibfnamefont {Andras}\ \bibnamefont {Kis}}, \
  and\ \bibinfo {author} {\bibfnamefont {Atac}\ \bibnamefont
  {Imamo{\u{g}}lu}},\ }\bibfield  {title} {\enquote {\bibinfo {title} {{Valley
  Zeeman effect in elementary optical excitations of monolayer WSe$_2$}},}\
  }\href@noop {} {\bibfield  {journal} {\bibinfo  {journal} {Nature Physics}\
  }\textbf {\bibinfo {volume} {11}},\ \bibinfo {pages} {141--147} (\bibinfo
  {year} {2015})}\BibitemShut {NoStop}%
\bibitem [{\citenamefont {F{\"o}rste}\ \emph {et~al.}(2020)\citenamefont
  {F{\"o}rste}, \citenamefont {Tepliakov}, \citenamefont {Kruchinin},
  \citenamefont {Lindlau}, \citenamefont {Funk}, \citenamefont {F{\"o}rg},
  \citenamefont {Watanabe}, \citenamefont {Taniguchi}, \citenamefont
  {Baimuratov},\ and\ \citenamefont {H{\"o}gele}}]{forste2020exciton}%
  \BibitemOpen
  \bibfield  {author} {\bibinfo {author} {\bibfnamefont {Jonathan}\
  \bibnamefont {F{\"o}rste}}, \bibinfo {author} {\bibfnamefont {Nikita~V}\
  \bibnamefont {Tepliakov}}, \bibinfo {author} {\bibfnamefont {Stanislav~Yu}\
  \bibnamefont {Kruchinin}}, \bibinfo {author} {\bibfnamefont {Jessica}\
  \bibnamefont {Lindlau}}, \bibinfo {author} {\bibfnamefont {Victor}\
  \bibnamefont {Funk}}, \bibinfo {author} {\bibfnamefont {Michael}\
  \bibnamefont {F{\"o}rg}}, \bibinfo {author} {\bibfnamefont {Kenji}\
  \bibnamefont {Watanabe}}, \bibinfo {author} {\bibfnamefont {Takashi}\
  \bibnamefont {Taniguchi}}, \bibinfo {author} {\bibfnamefont {Anvar~S}\
  \bibnamefont {Baimuratov}}, \ and\ \bibinfo {author} {\bibfnamefont
  {Alexander}\ \bibnamefont {H{\"o}gele}},\ }\bibfield  {title} {\enquote
  {\bibinfo {title} {{Exciton g-factors in monolayer and bilayer WSe$_2$ from
  experiment and theory}},}\ }\href@noop {} {\bibfield  {journal} {\bibinfo
  {journal} {{Nature Communications}}\ }\textbf {\bibinfo {volume} {11}},\
  \bibinfo {pages} {4539} (\bibinfo {year} {2020})}\BibitemShut {NoStop}%
\bibitem [{\citenamefont {Slagle}\ and\ \citenamefont
  {Fu}(2020)}]{slagle2020charge}%
  \BibitemOpen
  \bibfield  {author} {\bibinfo {author} {\bibfnamefont {Kevin}\ \bibnamefont
  {Slagle}}\ and\ \bibinfo {author} {\bibfnamefont {Liang}\ \bibnamefont
  {Fu}},\ }\bibfield  {title} {\enquote {\bibinfo {title} {Charge transfer
  excitations, pair density waves, and superconductivity in moir{\'e}
  materials},}\ }\href@noop {} {\bibfield  {journal} {\bibinfo  {journal}
  {Physical Review B}\ }\textbf {\bibinfo {volume} {102}},\ \bibinfo {pages}
  {235423} (\bibinfo {year} {2020})}\BibitemShut {NoStop}%
\bibitem [{\citenamefont {Dalal}\ and\ \citenamefont
  {Ruhman}(2021)}]{dalal2021orbitally}%
  \BibitemOpen
  \bibfield  {author} {\bibinfo {author} {\bibfnamefont {Amir}\ \bibnamefont
  {Dalal}}\ and\ \bibinfo {author} {\bibfnamefont {Jonathan}\ \bibnamefont
  {Ruhman}},\ }\bibfield  {title} {\enquote {\bibinfo {title} {{Orbitally
  selective Mott phase in electron-doped twisted transition
  metal-dichalcogenides: A possible realization of the Kondo lattice model}},}\
  }\href@noop {} {\bibfield  {journal} {\bibinfo  {journal} {Physical Review
  Research}\ }\textbf {\bibinfo {volume} {3}},\ \bibinfo {pages} {043173}
  (\bibinfo {year} {2021})}\BibitemShut {NoStop}%
\bibitem [{\citenamefont {Kumar}\ \emph {et~al.}(2022)\citenamefont {Kumar},
  \citenamefont {Hu}, \citenamefont {MacDonald},\ and\ \citenamefont
  {Potter}}]{kumar2022gate}%
  \BibitemOpen
  \bibfield  {author} {\bibinfo {author} {\bibfnamefont {Ajesh}\ \bibnamefont
  {Kumar}}, \bibinfo {author} {\bibfnamefont {Nai~Chao}\ \bibnamefont {Hu}},
  \bibinfo {author} {\bibfnamefont {Allan~H}\ \bibnamefont {MacDonald}}, \ and\
  \bibinfo {author} {\bibfnamefont {Andrew~C}\ \bibnamefont {Potter}},\
  }\bibfield  {title} {\enquote {\bibinfo {title} {{Gate-tunable heavy fermion
  quantum criticality in a moir{\'e} Kondo lattice}},}\ }\href@noop {}
  {\bibfield  {journal} {\bibinfo  {journal} {Physical Review B}\ }\textbf
  {\bibinfo {volume} {106}},\ \bibinfo {pages} {L041116} (\bibinfo {year}
  {2022})}\BibitemShut {NoStop}%
\bibitem [{\citenamefont {Raja}\ \emph {et~al.}(2017)\citenamefont {Raja},
  \citenamefont {Chaves}, \citenamefont {Yu}, \citenamefont {Arefe},
  \citenamefont {Hill}, \citenamefont {Rigosi}, \citenamefont {Berkelbach},
  \citenamefont {Nagler}, \citenamefont {Sch{\"u}ller}, \citenamefont {Korn}
  \emph {et~al.}}]{raja2017coulomb}%
  \BibitemOpen
  \bibfield  {author} {\bibinfo {author} {\bibfnamefont {Archana}\ \bibnamefont
  {Raja}}, \bibinfo {author} {\bibfnamefont {Andrey}\ \bibnamefont {Chaves}},
  \bibinfo {author} {\bibfnamefont {Jaeeun}\ \bibnamefont {Yu}}, \bibinfo
  {author} {\bibfnamefont {Ghidewon}\ \bibnamefont {Arefe}}, \bibinfo {author}
  {\bibfnamefont {Heather~M}\ \bibnamefont {Hill}}, \bibinfo {author}
  {\bibfnamefont {Albert~F}\ \bibnamefont {Rigosi}}, \bibinfo {author}
  {\bibfnamefont {Timothy~C}\ \bibnamefont {Berkelbach}}, \bibinfo {author}
  {\bibfnamefont {Philipp}\ \bibnamefont {Nagler}}, \bibinfo {author}
  {\bibfnamefont {Christian}\ \bibnamefont {Sch{\"u}ller}}, \bibinfo {author}
  {\bibfnamefont {Tobias}\ \bibnamefont {Korn}},  \emph {et~al.},\ }\bibfield
  {title} {\enquote {\bibinfo {title} {Coulomb engineering of the bandgap and
  excitons in two-dimensional materials},}\ }\href@noop {} {\bibfield
  {journal} {\bibinfo  {journal} {{Nature Communications}}\ }\textbf {\bibinfo
  {volume} {8}},\ \bibinfo {pages} {15251} (\bibinfo {year}
  {2017})}\BibitemShut {NoStop}%
\bibitem [{\citenamefont {Chen}\ \emph {et~al.}(2022)\citenamefont {Chen},
  \citenamefont {Lian}, \citenamefont {Huang}, \citenamefont {Su},
  \citenamefont {Rashetnia}, \citenamefont {Ma}, \citenamefont {Yan},
  \citenamefont {Blei}, \citenamefont {Xiang}, \citenamefont {Taniguchi} \emph
  {et~al.}}]{chen2022excitonic}%
  \BibitemOpen
  \bibfield  {author} {\bibinfo {author} {\bibfnamefont {Dongxue}\ \bibnamefont
  {Chen}}, \bibinfo {author} {\bibfnamefont {Zhen}\ \bibnamefont {Lian}},
  \bibinfo {author} {\bibfnamefont {Xiong}\ \bibnamefont {Huang}}, \bibinfo
  {author} {\bibfnamefont {Ying}\ \bibnamefont {Su}}, \bibinfo {author}
  {\bibfnamefont {Mina}\ \bibnamefont {Rashetnia}}, \bibinfo {author}
  {\bibfnamefont {Lei}\ \bibnamefont {Ma}}, \bibinfo {author} {\bibfnamefont
  {Li}~\bibnamefont {Yan}}, \bibinfo {author} {\bibfnamefont {Mark}\
  \bibnamefont {Blei}}, \bibinfo {author} {\bibfnamefont {Li}~\bibnamefont
  {Xiang}}, \bibinfo {author} {\bibfnamefont {Takashi}\ \bibnamefont
  {Taniguchi}},  \emph {et~al.},\ }\bibfield  {title} {\enquote {\bibinfo
  {title} {{Excitonic insulator in a heterojunction moir{\'e} superlattice}},}\
  }\href@noop {} {\bibfield  {journal} {\bibinfo  {journal} {Nature Physics}\
  }\textbf {\bibinfo {volume} {18}},\ \bibinfo {pages} {1171--1176} (\bibinfo
  {year} {2022})}\BibitemShut {NoStop}%
\bibitem [{\citenamefont {Sidler}\ \emph {et~al.}(2017)\citenamefont {Sidler},
  \citenamefont {Back}, \citenamefont {Cotlet}, \citenamefont {Srivastava},
  \citenamefont {Fink}, \citenamefont {Kroner}, \citenamefont {Demler},\ and\
  \citenamefont {Imamoglu}}]{sidler2017fermi}%
  \BibitemOpen
  \bibfield  {author} {\bibinfo {author} {\bibfnamefont {Meinrad}\ \bibnamefont
  {Sidler}}, \bibinfo {author} {\bibfnamefont {Patrick}\ \bibnamefont {Back}},
  \bibinfo {author} {\bibfnamefont {Ovidiu}\ \bibnamefont {Cotlet}}, \bibinfo
  {author} {\bibfnamefont {Ajit}\ \bibnamefont {Srivastava}}, \bibinfo {author}
  {\bibfnamefont {Thomas}\ \bibnamefont {Fink}}, \bibinfo {author}
  {\bibfnamefont {Martin}\ \bibnamefont {Kroner}}, \bibinfo {author}
  {\bibfnamefont {Eugene}\ \bibnamefont {Demler}}, \ and\ \bibinfo {author}
  {\bibfnamefont {Atac}\ \bibnamefont {Imamoglu}},\ }\bibfield  {title}
  {\enquote {\bibinfo {title} {Fermi polaron-polaritons in charge-tunable
  atomically thin semiconductors},}\ }\href@noop {} {\bibfield  {journal}
  {\bibinfo  {journal} {Nature Physics}\ }\textbf {\bibinfo {volume} {13}},\
  \bibinfo {pages} {255--261} (\bibinfo {year} {2017})}\BibitemShut {NoStop}%
\bibitem [{\citenamefont {Efimkin}\ and\ \citenamefont
  {MacDonald}(2017)}]{efimkin2017many}%
  \BibitemOpen
  \bibfield  {author} {\bibinfo {author} {\bibfnamefont {Dmitry~K}\
  \bibnamefont {Efimkin}}\ and\ \bibinfo {author} {\bibfnamefont {Allan~H}\
  \bibnamefont {MacDonald}},\ }\bibfield  {title} {\enquote {\bibinfo {title}
  {Many-body theory of trion absorption features in two-dimensional
  semiconductors},}\ }\href@noop {} {\bibfield  {journal} {\bibinfo  {journal}
  {Physical Review B}\ }\textbf {\bibinfo {volume} {95}},\ \bibinfo {pages}
  {035417} (\bibinfo {year} {2017})}\BibitemShut {NoStop}%
\bibitem [{\citenamefont {Fey}\ \emph {et~al.}(2020)\citenamefont {Fey},
  \citenamefont {Schmelcher}, \citenamefont {Imamoglu},\ and\ \citenamefont
  {Schmidt}}]{fey2020theory}%
  \BibitemOpen
  \bibfield  {author} {\bibinfo {author} {\bibfnamefont {Christian}\
  \bibnamefont {Fey}}, \bibinfo {author} {\bibfnamefont {Peter}\ \bibnamefont
  {Schmelcher}}, \bibinfo {author} {\bibfnamefont {Atac}\ \bibnamefont
  {Imamoglu}}, \ and\ \bibinfo {author} {\bibfnamefont {Richard}\ \bibnamefont
  {Schmidt}},\ }\bibfield  {title} {\enquote {\bibinfo {title} {{Theory of
  exciton-electron scattering in atomically thin semiconductors}},}\
  }\href@noop {} {\bibfield  {journal} {\bibinfo  {journal} {{Physical Review
  B}}\ }\textbf {\bibinfo {volume} {101}},\ \bibinfo {pages} {195417} (\bibinfo
  {year} {2020})}\BibitemShut {NoStop}%
\bibitem [{\citenamefont {Wang}\ \emph {et~al.}(2017)\citenamefont {Wang},
  \citenamefont {Zhao}, \citenamefont {Mak},\ and\ \citenamefont
  {Shan}}]{wang2017probing}%
  \BibitemOpen
  \bibfield  {author} {\bibinfo {author} {\bibfnamefont {Zefang}\ \bibnamefont
  {Wang}}, \bibinfo {author} {\bibfnamefont {Liang}\ \bibnamefont {Zhao}},
  \bibinfo {author} {\bibfnamefont {Kin~Fai}\ \bibnamefont {Mak}}, \ and\
  \bibinfo {author} {\bibfnamefont {Jie}\ \bibnamefont {Shan}},\ }\bibfield
  {title} {\enquote {\bibinfo {title} {Probing the spin-polarized electronic
  band structure in monolayer transition metal dichalcogenides by optical
  spectroscopy},}\ }\href@noop {} {\bibfield  {journal} {\bibinfo  {journal}
  {Nano letters}\ }\textbf {\bibinfo {volume} {17}},\ \bibinfo {pages}
  {740--746} (\bibinfo {year} {2017})}\BibitemShut {NoStop}%
\bibitem [{\citenamefont {Gao}\ \emph {et~al.}(2016)\citenamefont {Gao},
  \citenamefont {Liang}, \citenamefont {Spataru},\ and\ \citenamefont
  {Yang}}]{gao2016dynamical}%
  \BibitemOpen
  \bibfield  {author} {\bibinfo {author} {\bibfnamefont {Shiyuan}\ \bibnamefont
  {Gao}}, \bibinfo {author} {\bibfnamefont {Yufeng}\ \bibnamefont {Liang}},
  \bibinfo {author} {\bibfnamefont {Catalin~D}\ \bibnamefont {Spataru}}, \ and\
  \bibinfo {author} {\bibfnamefont {Li}~\bibnamefont {Yang}},\ }\bibfield
  {title} {\enquote {\bibinfo {title} {Dynamical excitonic effects in doped
  two-dimensional semiconductors},}\ }\href@noop {} {\bibfield  {journal}
  {\bibinfo  {journal} {{Nano Letters}}\ }\textbf {\bibinfo {volume} {16}},\
  \bibinfo {pages} {5568--5573} (\bibinfo {year} {2016})}\BibitemShut {NoStop}%
\bibitem [{\citenamefont {Yao}\ \emph {et~al.}(2017)\citenamefont {Yao},
  \citenamefont {Yan}, \citenamefont {Kahn}, \citenamefont {Suslu},
  \citenamefont {Liang}, \citenamefont {Barnard}, \citenamefont {Tongay},
  \citenamefont {Zettl}, \citenamefont {Borys},\ and\ \citenamefont
  {Schuck}}]{yao2017optically}%
  \BibitemOpen
  \bibfield  {author} {\bibinfo {author} {\bibfnamefont {Kaiyuan}\ \bibnamefont
  {Yao}}, \bibinfo {author} {\bibfnamefont {Aiming}\ \bibnamefont {Yan}},
  \bibinfo {author} {\bibfnamefont {Salman}\ \bibnamefont {Kahn}}, \bibinfo
  {author} {\bibfnamefont {Aslihan}\ \bibnamefont {Suslu}}, \bibinfo {author}
  {\bibfnamefont {Yufeng}\ \bibnamefont {Liang}}, \bibinfo {author}
  {\bibfnamefont {Edward~S}\ \bibnamefont {Barnard}}, \bibinfo {author}
  {\bibfnamefont {Sefaattin}\ \bibnamefont {Tongay}}, \bibinfo {author}
  {\bibfnamefont {Alex}\ \bibnamefont {Zettl}}, \bibinfo {author}
  {\bibfnamefont {Nicholas~J}\ \bibnamefont {Borys}}, \ and\ \bibinfo {author}
  {\bibfnamefont {P~James}\ \bibnamefont {Schuck}},\ }\bibfield  {title}
  {\enquote {\bibinfo {title} {{Optically discriminating carrier-induced
  quasiparticle band gap and exciton energy renormalization in monolayer
  MoS$_2$}},}\ }\href@noop {} {\bibfield  {journal} {\bibinfo  {journal}
  {Physical Review Letters}\ }\textbf {\bibinfo {volume} {119}},\ \bibinfo
  {pages} {087401} (\bibinfo {year} {2017})}\BibitemShut {NoStop}%
\bibitem [{\citenamefont {Jauregui}\ \emph {et~al.}(2019)\citenamefont
  {Jauregui}, \citenamefont {Joe}, \citenamefont {Pistunova}, \citenamefont
  {Wild}, \citenamefont {High}, \citenamefont {Zhou}, \citenamefont {Scuri},
  \citenamefont {De~Greve}, \citenamefont {Sushko}, \citenamefont {Yu} \emph
  {et~al.}}]{jauregui2019electrical}%
  \BibitemOpen
  \bibfield  {author} {\bibinfo {author} {\bibfnamefont {Luis~A}\ \bibnamefont
  {Jauregui}}, \bibinfo {author} {\bibfnamefont {Andrew~Y}\ \bibnamefont
  {Joe}}, \bibinfo {author} {\bibfnamefont {Kateryna}\ \bibnamefont
  {Pistunova}}, \bibinfo {author} {\bibfnamefont {Dominik~S}\ \bibnamefont
  {Wild}}, \bibinfo {author} {\bibfnamefont {Alexander~A}\ \bibnamefont
  {High}}, \bibinfo {author} {\bibfnamefont {You}\ \bibnamefont {Zhou}},
  \bibinfo {author} {\bibfnamefont {Giovanni}\ \bibnamefont {Scuri}}, \bibinfo
  {author} {\bibfnamefont {Kristiaan}\ \bibnamefont {De~Greve}}, \bibinfo
  {author} {\bibfnamefont {Andrey}\ \bibnamefont {Sushko}}, \bibinfo {author}
  {\bibfnamefont {Che-Hang}\ \bibnamefont {Yu}},  \emph {et~al.},\ }\bibfield
  {title} {\enquote {\bibinfo {title} {Electrical control of interlayer exciton
  dynamics in atomically thin heterostructures},}\ }\href@noop {} {\bibfield
  {journal} {\bibinfo  {journal} {Science}\ }\textbf {\bibinfo {volume}
  {366}},\ \bibinfo {pages} {870--875} (\bibinfo {year} {2019})}\BibitemShut
  {NoStop}%
\bibitem [{\citenamefont {Brixner}(1962)}]{brixner1962preparation}%
  \BibitemOpen
  \bibfield  {author} {\bibinfo {author} {\bibfnamefont {Lothar~H}\
  \bibnamefont {Brixner}},\ }\bibfield  {title} {\enquote {\bibinfo {title}
  {Preparation and properties of the single crystalline ab2-type selenides and
  tellurides of niobium, tantalum, molybdenum and tungsten},}\ }\href@noop {}
  {\bibfield  {journal} {\bibinfo  {journal} {{Journal of Inorganic and Nuclear
  Chemistry}}\ }\textbf {\bibinfo {volume} {24}},\ \bibinfo {pages} {257--263}
  (\bibinfo {year} {1962})}\BibitemShut {NoStop}%
\bibitem [{\citenamefont {Laturia}\ \emph {et~al.}(2018)\citenamefont
  {Laturia}, \citenamefont {Van~de Put},\ and\ \citenamefont
  {Vandenberghe}}]{laturia2018dielectric}%
  \BibitemOpen
  \bibfield  {author} {\bibinfo {author} {\bibfnamefont {Akash}\ \bibnamefont
  {Laturia}}, \bibinfo {author} {\bibfnamefont {Maarten~L}\ \bibnamefont
  {Van~de Put}}, \ and\ \bibinfo {author} {\bibfnamefont {William~G}\
  \bibnamefont {Vandenberghe}},\ }\bibfield  {title} {\enquote {\bibinfo
  {title} {Dielectric properties of hexagonal boron nitride and transition
  metal dichalcogenides: from monolayer to bulk},}\ }\href@noop {} {\bibfield
  {journal} {\bibinfo  {journal} {npj 2D Materials and Applications}\ }\textbf
  {\bibinfo {volume} {2}} (\bibinfo {year} {2018})}\BibitemShut {NoStop}%
\bibitem [{\citenamefont {Baek}\ \emph {et~al.}(2020)\citenamefont {Baek},
  \citenamefont {Brotons-Gisbert}, \citenamefont {Koong}, \citenamefont
  {Campbell}, \citenamefont {Rambach}, \citenamefont {Watanabe}, \citenamefont
  {Taniguchi},\ and\ \citenamefont {Gerardot}}]{baek2020highly}%
  \BibitemOpen
  \bibfield  {author} {\bibinfo {author} {\bibfnamefont {H}~\bibnamefont
  {Baek}}, \bibinfo {author} {\bibfnamefont {M}~\bibnamefont
  {Brotons-Gisbert}}, \bibinfo {author} {\bibfnamefont {ZX}~\bibnamefont
  {Koong}}, \bibinfo {author} {\bibfnamefont {A}~\bibnamefont {Campbell}},
  \bibinfo {author} {\bibfnamefont {M}~\bibnamefont {Rambach}}, \bibinfo
  {author} {\bibfnamefont {K}~\bibnamefont {Watanabe}}, \bibinfo {author}
  {\bibfnamefont {T}~\bibnamefont {Taniguchi}}, \ and\ \bibinfo {author}
  {\bibfnamefont {BD}~\bibnamefont {Gerardot}},\ }\bibfield  {title} {\enquote
  {\bibinfo {title} {Highly energy-tunable quantum light from moir{\'e}-trapped
  excitons},}\ }\href@noop {} {\bibfield  {journal} {\bibinfo  {journal}
  {Science Advances}\ }\textbf {\bibinfo {volume} {6}},\ \bibinfo {pages}
  {eaba8526} (\bibinfo {year} {2020})}\BibitemShut {NoStop}%
\bibitem [{\citenamefont {Brotons-Gisbert}\ \emph {et~al.}(2021)\citenamefont
  {Brotons-Gisbert}, \citenamefont {Baek}, \citenamefont {Campbell},
  \citenamefont {Watanabe}, \citenamefont {Taniguchi},\ and\ \citenamefont
  {Gerardot}}]{brotons2021moir}%
  \BibitemOpen
  \bibfield  {author} {\bibinfo {author} {\bibfnamefont {Mauro}\ \bibnamefont
  {Brotons-Gisbert}}, \bibinfo {author} {\bibfnamefont {Hyeonjun}\ \bibnamefont
  {Baek}}, \bibinfo {author} {\bibfnamefont {Aidan}\ \bibnamefont {Campbell}},
  \bibinfo {author} {\bibfnamefont {Kenji}\ \bibnamefont {Watanabe}}, \bibinfo
  {author} {\bibfnamefont {Takashi}\ \bibnamefont {Taniguchi}}, \ and\ \bibinfo
  {author} {\bibfnamefont {Brian~D}\ \bibnamefont {Gerardot}},\ }\bibfield
  {title} {\enquote {\bibinfo {title} {{Moir{\'e}-trapped interlayer trions in
  a charge-tunable WSe$_2$/MoSe$_2 $ heterobilayer}},}\ }\href@noop {}
  {\bibfield  {journal} {\bibinfo  {journal} {{Physical Review X}}\ }\textbf
  {\bibinfo {volume} {11}},\ \bibinfo {pages} {031033} (\bibinfo {year}
  {2021})}\BibitemShut {NoStop}%
\bibitem [{\citenamefont {Thompson}\ \emph {et~al.}(2022)\citenamefont
  {Thompson}, \citenamefont {Aktulga}, \citenamefont {Berger}, \citenamefont
  {Bolintineanu}, \citenamefont {Brown}, \citenamefont {Crozier}, \citenamefont
  {in~'t Veld}, \citenamefont {Kohlmeyer}, \citenamefont {Moore}, \citenamefont
  {Nguyen}, \citenamefont {Shan}, \citenamefont {Stevens}, \citenamefont
  {Tranchida}, \citenamefont {Trott},\ and\ \citenamefont {Plimpton}}]{LAMMPS}%
  \BibitemOpen
  \bibfield  {author} {\bibinfo {author} {\bibfnamefont {A.~P.}\ \bibnamefont
  {Thompson}}, \bibinfo {author} {\bibfnamefont {H.~M.}\ \bibnamefont
  {Aktulga}}, \bibinfo {author} {\bibfnamefont {R.}~\bibnamefont {Berger}},
  \bibinfo {author} {\bibfnamefont {D.~S.}\ \bibnamefont {Bolintineanu}},
  \bibinfo {author} {\bibfnamefont {W.~M.}\ \bibnamefont {Brown}}, \bibinfo
  {author} {\bibfnamefont {P.~S.}\ \bibnamefont {Crozier}}, \bibinfo {author}
  {\bibfnamefont {P.~J.}\ \bibnamefont {in~'t Veld}}, \bibinfo {author}
  {\bibfnamefont {A.}~\bibnamefont {Kohlmeyer}}, \bibinfo {author}
  {\bibfnamefont {S.~G.}\ \bibnamefont {Moore}}, \bibinfo {author}
  {\bibfnamefont {T.~D.}\ \bibnamefont {Nguyen}}, \bibinfo {author}
  {\bibfnamefont {R.}~\bibnamefont {Shan}}, \bibinfo {author} {\bibfnamefont
  {M.~J.}\ \bibnamefont {Stevens}}, \bibinfo {author} {\bibfnamefont
  {J.}~\bibnamefont {Tranchida}}, \bibinfo {author} {\bibfnamefont
  {C.}~\bibnamefont {Trott}}, \ and\ \bibinfo {author} {\bibfnamefont {S.~J.}\
  \bibnamefont {Plimpton}},\ }\bibfield  {title} {\enquote {\bibinfo {title}
  {{LAMMPS} - a flexible simulation tool for particle-based materials modeling
  at the atomic, meso, and continuum scales},}\ }\href {\doibase
  10.1016/j.cpc.2021.108171} {\bibfield  {journal} {\bibinfo  {journal} {Comp.
  Phys. Comm.}\ }\textbf {\bibinfo {volume} {271}},\ \bibinfo {pages} {108171}
  (\bibinfo {year} {2022})}\BibitemShut {NoStop}%
\bibitem [{\citenamefont {Soler}\ \emph {et~al.}(2002)\citenamefont {Soler},
  \citenamefont {Artacho}, \citenamefont {Gale}, \citenamefont {Garc\'ia},
  \citenamefont {Junquera}, \citenamefont {Ordej\'on},\ and\ \citenamefont
  {S\'anchez-Portal}}]{Jose_M_Soler_2002}%
  \BibitemOpen
  \bibfield  {author} {\bibinfo {author} {\bibfnamefont {Jos\'e~M}\
  \bibnamefont {Soler}}, \bibinfo {author} {\bibfnamefont {Emilio}\
  \bibnamefont {Artacho}}, \bibinfo {author} {\bibfnamefont {Julian~D}\
  \bibnamefont {Gale}}, \bibinfo {author} {\bibfnamefont {Alberto}\
  \bibnamefont {Garc\'ia}}, \bibinfo {author} {\bibfnamefont {Javier}\
  \bibnamefont {Junquera}}, \bibinfo {author} {\bibfnamefont {Pablo}\
  \bibnamefont {Ordej\'on}}, \ and\ \bibinfo {author} {\bibfnamefont {Daniel}\
  \bibnamefont {S\'anchez-Portal}},\ }\bibfield  {title} {\enquote {\bibinfo
  {title} {The siesta method for ab initio order-n materials simulation},}\
  }\href {\doibase 10.1088/0953-8984/14/11/302} {\bibfield  {journal} {\bibinfo
   {journal} {Journal of Physics: Condensed Matter}\ }\textbf {\bibinfo
  {volume} {14}},\ \bibinfo {pages} {2745} (\bibinfo {year}
  {2002})}\BibitemShut {NoStop}%
\bibitem [{\citenamefont {Kohn}\ and\ \citenamefont {Sham}(1965)}]{LDA}%
  \BibitemOpen
  \bibfield  {author} {\bibinfo {author} {\bibfnamefont {W.}~\bibnamefont
  {Kohn}}\ and\ \bibinfo {author} {\bibfnamefont {L.~J.}\ \bibnamefont
  {Sham}},\ }\bibfield  {title} {\enquote {\bibinfo {title} {Self-consistent
  equations including exchange and correlation effects},}\ }\href {\doibase
  10.1103/PhysRev.140.A1133} {\bibfield  {journal} {\bibinfo  {journal} {Phys.
  Rev.}\ }\textbf {\bibinfo {volume} {140}},\ \bibinfo {pages} {A1133--A1138}
  (\bibinfo {year} {1965})}\BibitemShut {NoStop}%
\bibitem [{\citenamefont {Fern\'andez-Seivane}\ \emph
  {et~al.}(2006)\citenamefont {Fern\'andez-Seivane}, \citenamefont {Oliveira},
  \citenamefont {Sanvito},\ and\ \citenamefont
  {Ferrer}}]{Fernandez_Seivane_2006}%
  \BibitemOpen
  \bibfield  {author} {\bibinfo {author} {\bibfnamefont {L}~\bibnamefont
  {Fern\'andez-Seivane}}, \bibinfo {author} {\bibfnamefont {M~A}\ \bibnamefont
  {Oliveira}}, \bibinfo {author} {\bibfnamefont {S}~\bibnamefont {Sanvito}}, \
  and\ \bibinfo {author} {\bibfnamefont {J}~\bibnamefont {Ferrer}},\ }\bibfield
   {title} {\enquote {\bibinfo {title} {On-site approximation for spin–orbit
  coupling in linear combination of atomic orbitals density functional
  methods},}\ }\href {\doibase 10.1088/0953-8984/18/34/012} {\bibfield
  {journal} {\bibinfo  {journal} {Journal of Physics: Condensed Matter}\
  }\textbf {\bibinfo {volume} {18}},\ \bibinfo {pages} {7999} (\bibinfo {year}
  {2006})}\BibitemShut {NoStop}%
\bibitem [{\citenamefont {Troullier}\ and\ \citenamefont
  {Martins}(1991)}]{TroulliersMartins1991}%
  \BibitemOpen
  \bibfield  {author} {\bibinfo {author} {\bibfnamefont {N.}~\bibnamefont
  {Troullier}}\ and\ \bibinfo {author} {\bibfnamefont {Jos\'e~Lu\'{\i}s}\
  \bibnamefont {Martins}},\ }\bibfield  {title} {\enquote {\bibinfo {title}
  {Efficient pseudopotentials for plane-wave calculations},}\ }\href {\doibase
  10.1103/PhysRevB.43.1993} {\bibfield  {journal} {\bibinfo  {journal} {Phys.
  Rev. B}\ }\textbf {\bibinfo {volume} {43}},\ \bibinfo {pages} {1993--2006}
  (\bibinfo {year} {1991})}\BibitemShut {NoStop}%
\bibitem [{\citenamefont {Papior}\ \emph {et~al.}(2017)\citenamefont {Papior},
  \citenamefont {Lorente}, \citenamefont {Frederiksen}, \citenamefont
  {García},\ and\ \citenamefont {Brandbyge}}]{PAPIOR20178}%
  \BibitemOpen
  \bibfield  {author} {\bibinfo {author} {\bibfnamefont {Nick}\ \bibnamefont
  {Papior}}, \bibinfo {author} {\bibfnamefont {Nicolás}\ \bibnamefont
  {Lorente}}, \bibinfo {author} {\bibfnamefont {Thomas}\ \bibnamefont
  {Frederiksen}}, \bibinfo {author} {\bibfnamefont {Alberto}\ \bibnamefont
  {García}}, \ and\ \bibinfo {author} {\bibfnamefont {Mads}\ \bibnamefont
  {Brandbyge}},\ }\bibfield  {title} {\enquote {\bibinfo {title} {Improvements
  on non-equilibrium and transport green function techniques: The
  next-generation transiesta},}\ }\href {\doibase
  https://doi.org/10.1016/j.cpc.2016.09.022} {\bibfield  {journal} {\bibinfo
  {journal} {Computer Physics Communications}\ }\textbf {\bibinfo {volume}
  {212}},\ \bibinfo {pages} {8--24} (\bibinfo {year} {2017})}\BibitemShut
  {NoStop}%
\end{thebibliography}%

\section{Data Availability}

The datasets generated and analysed during the current study are available from the corresponding author on reasonable request.

\section{Acknowledgements}

We thank Jonathan Ruhman for insightful discussions. This work was supported by the EPSRC (grant nos. EP/P029892/1 and EP/L015110/1), and the ERC (grant no. 725920). V.V and J.L. acknowledge funding from the EPSRC (grant no. EP/S025324/1). This work used the ARCHER2 UK National Supercomputing Service via J.L.’s membership of the HEC
Materials Chemistry Consortium of UK, which is funded
by EPSRC (EP/L000202). This  project  has  received  funding  from  the  European  Union’s  Horizon  2020  research  and  innovation programme under the Marie Sk\l{}odowska-Curie grant agreement No. 101067977.
M.B.-G. is supported by a Royal Society University Research Fellowship. B.D.G. is supported by a Wolfson Merit Award from the Royal Society and a Chair in Emerging Technology from the Royal Academy of Engineering. K.W. and T.T. acknowledge support
from the Elemental Strategy Initiative conducted by the MEXT, Japan, grant no.
JPMXP0112101001, JSPS KAKENHI grant no. 19H05790 and JP20H00354. 

\section{Contributions}

A.C. performed the experimental measurements. A.C., M.B.G., and B.D.G. analysed the data. H.B. fabricated the sample. V.V. and J.L. performed the theoretical calculations. T.T. and J.W. grew the hBN crystals. B.D.G. and J.L. conceived and supervised the project. A.C., V.V., M.B.G, J.L. and B.D.G. and wrote the manuscript with input from all authors.

\end{document}